\documentclass{emulateapj}

\begin{document}
\title{Modeling Mid-Infrared Variability of Circumstellar Disks with Non-Axisymmetric Structure}
\author{Flaherty, K.M. \altaffilmark{1},Muzerolle, J. \altaffilmark{1,2}}
\email{kflaherty@as.arizona.edu}

\altaffiltext{1}{Steward Observatory, University of Arizona, Tucson, AZ 85721}
\altaffiltext{2}{Space Telescope Science Institute, 3700 San Martin Dr., Baltimore, MD 21218}

\begin{abstract}
Recent mid-infrared observations  of young stellar objects have found significant variations possibly indicative of changes in the structure of the circumstellar disk. Previous models of this variability have been restricted to axisymmetric perturbations in the disk. We consider simple models of a non-axisymmetric variation in the inner disk, such as a warp or a spiral wave. We find that the precession of these non-axisymmetric structures produce negligible flux variations but a change in the height of these structures can lead to significant changes in the mid-infrared flux. Applying these models to observations of the young stellar object LRLL 31 suggests that the observed variability could be explained by a warped inner disk with variable scale height. This suggests that some of the variability observed in young stellar objects could be explained by non-axisymmetric disturbances in the inner disk and this variability would be easily observable in future studies.

\end{abstract}


\section{Introduction}
Variability has been one of the prominent features of T Tauri stars since they were first observed \citep{joy45}. In the optical flux changes up to 0.5 magnitudes over a few days are common with most of the variability explained by hot and cold spots rotating across the surface of the star \citep{her94}. In a few cases, circumstellar disk variability has been inferred from stochastic or quasi-periodic optical occultation events. In the case of AA Tau a warped inner disk occults the star leading to periodic optical variations \citep{bou03}. For HH 30, putative structural changes in the disk may lead to the observed changes in the scattered light emission in as little as a few days \citep{wat07}. Variations in the scattered light images of HD 163296 are also attributed to a change in the structure of the disk \citep{wis08}. FU Ori outbursts, which show an orders of magnitude increase in the optical flux  followed by a slow decrease over decades, are associated with changes in the accretion rate of the disk brought about by thermal or gravitational instabilities \citep{har96}. EXor stars also show variability on month to yearly timescales probably related to changes in the accretion rate through the disk \citep{her07}.

Infrared observations provide a more direct way of looking for variability of the circumstellar disk. Emission at these wavelengths comes from dust near the surface of the central $\lesssim$10 AU of the disk and is sensitive to changes in the disk temperature and structure. The dust temperature is set by a combination of viscous energy release as disk gas accretes toward the central star and irradiation of the disk surface by light from the stellar photosphere and accretion shock. Variations in the near-infrared cannot always be explained simply by a change in the stellar flux, but a change in the structure of the circumstellar disk must also be included \citep{car01,eir02,skr96,alv08}. \citet{pla08} find variable disk emission at 3-8 \micron\ around the star WL 4, which also exhibits periodic near-infrared changes suggestive of influence by a binary companion. Variability as far out as 100\micron\ has been observed \citep{liu96,juh07,bar05} although the limited wavelength coverage and time sampling do not provide strong constraints on the physical cause. Some of this variation can be explained by changes in the structure of the inner disk. Variability of the 10\micron\ silicate feature, which is sensitive to grain properties at the surface of the inner few AU of the disk, has also been observed suggesting that the grain properties of disks are rapidly varying\citep{abr09,ske10}. 

Most disk models assume axisymmetry, and thus explore axisymmetric structure variations or asymmetric irradiation fields to explain the observed variability. \citet{muz09} suggest that a non-axisymmetric inner disk may explain the strange wavelength dependence of the variations seen in LRLL 31, a transition disk in the young cluster IC 348. In this paper we further explore the possibility that these mid-infrared variations could be caused by a non-axisymmetric structure in the inner disk. First we consider three simple models for a non-axisymmetric structure: a warp in the middle of the disk, a warp at the inner edge of the disk and a spiral wave in the disk. We examine how the SED of these disks varies as the structures precesses/corotates around the star and as the height of the structure changes with time.  We then apply these models to the variations in LRLL 31 as well as to other mid-infrared variable stars. 

\section{Models\label{models}}
\subsection{Motivation}

A number of authors have recently found strong wavelength dependent mid-infrared variability in both Herbig AeBe stars and T Tauri stars. \citet{juh07} observe 3-100\micron\ variability of SV Cep over a two year period. These variations can be up to a factor of two at 100\micron, and there appears to be a correlation between the 100\micron\ flux and the optical flux and a weak anti-correlation between the 3.6\micron\ flux and the optical flux. \citet{sit08} find significant variations in the 3-13\micron\ flux of the Herbig AeBe star HD 31648 over almost a quarter-century. \citet{hut94} find two stars, UX Ori and AK Sco, whose optical flux decreases as the 10\micron\ flux increases. \citet{mor09} find that 3-8\micron\ variability of 0.05-0.2 mag is common in pre-main sequence stars in the young cluster IC 1396A. \citet{muz09} find mid-infrared variations of up to 60\% in as little as one week in the young star LRLL 31. The IRS spectra show a clear wavelength dependence of the flux variations, where the 5-8\micron\ flux decreases and the 8-40\micron\ flux increases. This variation is especially interesting since LRLL 31 exhibits an SED typical of a transition disk, indicative of a deficit of small grains in the inner disk.

Typical sources of optical variability in pre-main sequence stars have difficulty explaining the observed mid-infrared variations in these stars. Cold spots on the star cannot produce a large enough variation in the mid-infrared, where the flux is dominated by the circumstellar disk. They would not be able to produce variability out as far as 100\micron, as has been seen in a number of stars. Hot spots, due to the accretion flow striking the surface of the star, produce large optical variations that may affect the infrared flux. As the star rotates, the hot spot acts as a flashlight heating a localized portion of the disk. If the disk is viewed close to edge-on then the asymmetric heating can produce noticeable variations in the infrared \citep{mor09}. However, it is difficult for this model to explain some of the larger variations and different wavelength dependences seen in many young stellar objects.

\citet{juh07} and \citet{sit08} use detailed radiative transfer models of a varying height at the inner edge of the disk to reproduce their observations. Increasing the scale height raises the near-infrared flux while shadowing the outer disk, and occulting the star if the system is close to edge on. The long-wavelength variation is difficult to explain with in-situ changes in the structure of the disk because the dynamical timescale is years to decades longer than the observations, thus models with variations in the inner disk that shadow the outer disk are used. Due to the restrictions of the radiative transfer code, the perturbations were assumed to be axisymmetric within the disk. \citet{muz09} suggest that the variation in LRLL 31 may be a warp, spiral wave or some other non-axisymmetric structure in the inner disk. Some of the observations of LRLL 31 are difficult to explain with a puffed inner rim and other configurations were considered. Here we follow up on that possibility and develop a simple model of such structures. When examining these non-axisymmetric models we consider two time-dependent effects: azimuthal motion such as precession and corotation, and variations in the scale height of the perturbation. Precession is a natural consequence of the torques that originally warped the disk while a variable scale height may be related to the dynamical process that causes a warp in the first place. Both of these are possible sources of variability for a non-axisymmetric structure.

\subsection{Deriving the SED}
Disks are generally modeled as axisymmetric flared disks that are heated by stellar irradiation \citep{chi97} or a combination of stellar irradiation and viscous dissipation from accretion \citep{dal06}. But the disks can become disturbed from this initial state by the gravitational influence of a binary companion or passing star \citep{art94,lar97} or an instability within the disk \citep{wat08}. This can lead to the disk becoming warped, which will change the temperature structure and emission of the disk. To study how structural changes will effect the SED we consider three simple models (1) a disk with a middle warp, (2) a disk with an inner warp and (3) a disk with a spiral wave, shown in figure~\ref{drawing}. All of these disks are taken to be geometrically thin blackbody emitters. Flaring and an optically thin atmosphere will have a substantial effect on the SED but these simple models present a useful starting point. If our models were completely confined to the midplane the SED would have the simple form $\lambda F_{\lambda}\propto\lambda^{-4/3}$. How our models depart from the power law will depend on the location of the structural disturbance within the disk. To calculate the SED for our simple models we follow the derivation presented in \citet[hereafter TB96]{tb96}. TB96 lay out a procedure for estimating the temperature structure of a blackbody disk whose height, $h$, is an arbitrary function of $r$ and $\theta$ within the disk. The coordinate system is defined as centered on the star with $\theta=0$ along the line with the highest warp above the midplane. The flux intercepting a point $P(r,\theta)$ in the disk is given by:

\begin{eqnarray}
\textstyle
F_{\lambda}=2B_{\lambda}(T_{*})\int_{\delta_{min}}^{\delta_{max}}\gamma\sin(\delta) ( \int_{\varepsilon_{min}}^{\varepsilon_{max}}|A\cos\delta\\ \textstyle
+B\sin\delta\sin\varepsilon+C\sin\delta\cos\varepsilon|d\varepsilon ) d\delta\nonumber
\end{eqnarray}
where

\begin{equation}
\gamma=\frac1{\sqrt{1+(\partial h/\partial r)^2+(\partial h/\partial \theta)^2/r^2}}
\end{equation}

and 

\begin{mathletters}
\begin{eqnarray}
A=\left(\frac{\partial h}{\partial r}\right )\cos\beta-\sin\beta\\
B=\frac1{r}\left(\frac{\partial h}{\partial \theta}\right )\\
C=\left(\frac{\partial h}{\partial r}\right )\sin\beta+\cos\beta
\end{eqnarray}
\end{mathletters}

with $tan(\beta)=h(r,\theta)/r$. The variables $\delta$ and $\varepsilon$ are the polar and azimuthal coordinates of a point on the surface of the star with respect to the axis joining the polar axis to the center of the star. Converting to temperatures, and performing the integration over $\varepsilon$, the equation then becomes:

\begin{eqnarray}
T^4(r,\theta)=\frac2{\pi}T_*^4\int_{\delta_{min}}^{\delta_{max}}\gamma\sin\delta\{A[\varepsilon_{max}-\varepsilon_{min}]\cos\delta\\ \nonumber
+B[-\cos\varepsilon_{max}+\cos\varepsilon_{min}]\sin\delta\\ \nonumber
+C[\sin\varepsilon_{max}-\sin\varepsilon_{min}]\sin\delta\}d\delta\\ \nonumber
+\frac{3G\dot{M}M_*}{8\pi\sigma r^3}\left(1-\sqrt{\frac{R_*}{r}}\right )\nonumber
\end{eqnarray}

The last term in this equation is the energy from viscous dissipation, which for typical T Tauri accretion rates is much less than the heating from irradiation. We assume that the accretion rate onto the star is small enough that the accretion luminosity is negligible compared to the stellar luminosity and thus does not contribute to the heating of the disk. The process of evaluating the temperature of the disk reduces to a determination of $\delta_{max},\delta_{min},\varepsilon_{min},\varepsilon_{max}$. The three models will vary in their evaluation of $\delta_{max},\delta_{min},\varepsilon_{min},\varepsilon_{max}$ for each point $P(r,\theta)$ on the disk surface, which is discussed in more detail in the appendix.

From TB96, the flux emitted by the disk is:

\begin{equation}
F_{\nu,{\bf u}}= \int\int_{disk surface} B_{\nu}(T(r,\theta))dS{\bf n_d}\cdot{\bf u}
\end{equation}

In this case {\bf u} is the vector along the line of sight to the observer from the center of the star and ${\bf n_d}$ is the normal to the disk at the point $P(r,\theta)$. The vector ${\bf u}$ can be defined in terms of the azimuthal and polar angles to the line of sight, $\alpha$ and $i$ respectively. The disk surface extends from $r_{min}$, which we take to be $8R_*$, out to $R_{disk}$, typically 100 AU. The inner radius $8R_*$ is roughly equal to the dust destruction radius for a frontally illuminated disk with well-mixed small dust grains around a 0.5$M_{\odot}$ T Tauri star \citep{muz03}.

The area of the disk along the line of sight is given by

\begin{eqnarray}
\textstyle
dS {\bf n_d}\cdot{\bf u} = r[\left(\frac1{r}\frac{\partial h}{\partial \theta}\sin\theta-\frac{\partial h}{\partial r}\cos\theta\right)\cos\alpha\sin i\\ \nonumber \textstyle
-\left(\frac1{r}\frac{\partial h}{\partial\theta}\cos\theta
+\frac{\partial h}{\partial r}\sin\theta\right)\sin\alpha\sin i+\cos i]drd\theta
\end{eqnarray}

We also include the possibility of the star being occulted by the disk. These prescriptions are described in detail in the appendix. This is highly dependent on the viewing angle, which for the non-axisymmetric disks considered here depends on both inclination, $i$, and the azimuthal viewing angle, $\alpha$. The line of sight $\alpha=0$ corresponds to looking at the disk along the line $\theta=0$, while an observer along $i=0^{\circ}$ is viewing the disk face-on. TB96 do not examine the dependence on azimuthal angle, although their generic equations repeated above do include $\alpha$. In the appendix we demonstrate how this dependence has been included in the models. 

For our fiducial models we take the central source to be a G6 star with a radius of 2.9 $R_{\odot}$, a mass of 1.8 $M_{\odot}$ and an accretion rate of $\dot{M}=10^{-8}M_{\odot} yr^{-1}$.  These specific parameters are chosen in anticipation of modeling the disk around LRLL 31. As we will show later, the shape and strength of the variability does not depend strongly on the central star and the results presented in the next sections are applicable to a broad range of pre-main sequence stars.


\subsection{Middle Warp}
We first consider a warp similar to that originally discussed by TB96 whose functional form is:

 \begin{equation}
h(r,\theta)=gR_{warp}\left(\frac{r}{R_{warp}}\right)^n\cos(\theta)
\end{equation}

for $r\leqq R_{warp}$. The parameter $R_{warp}$ specifies the outer radius of the warp. TB96 take $R_{warp}$ to be equal to the radius of the disk, while here we consider situations where the disk extends beyond the warp, shown in figure 1a, in order to examine shadowing effects. Past $R_{warp}$ the disk is confined to the midplane and this flat passive extension can be shadowed by the warp, changing its temperature structure. Since the warp occurs in the middle of the disk we refer to it as a middle warp. We consider disks with $R_{warp}$ of a few AU or less in order to probe any effects on the 2 to 24\micron\ flux. The parameter $g$ sets the height of the warp, and is equivalent to the maximum value of $h/r$ within the disk.  According to \citet{tb93} a warp created by an equal mass binary will have $n=4$, $g=0.01$. Figure~\ref{warp1} shows the temperature structure of disks with $g=0.01,0.05$ and $R_{warp}=0.5$ AU. These warps demonstrate the shadowing of the outer disk and how it can change with radius. Immediately behind the warp (dark line beyond 0.5 AU) the disk is completely shadowed and is only heated by viscous dissipation. Further out the warp only blocks part of the star and the temperature approaches that of a flat passive disk. The height of the warp will change the temperature at the warp as well as the shadowing of the outer disk. Figure~\ref{warp1} compares the SED of this warped disk with a flat passive disk and a flat accretion disk. The short wavelength flux rises above a flat passive disk because of the high temperature on the side of the warp facing the star. The shadowing of the outer disk reduces the flux at long wavelengths so that it is less than a flat passive disk.

A warped disk will be driven to precess by the torques on the system which originally warped the disk. To mimic a precessing warped disk, we vary the azimuthal viewing angle, $\alpha$, of the observer. The method of TB96 needs to be modified to account for a changing azimuthal angle, and these changes are described in the appendix. In figure ~\ref{warp_prec} we show the precession of a warp with $g=0.05$ which extends to $R_{warp}=0.5$AU with a flat disk extending to 100 AU. As the disk precesses the infrared excess only changes by $\sim0.05$ dex. The flux change at short and long wavelengths is correlated. The observed wavelength dependence can be understood by considering the schematic middle warp drawn in figure~\ref{drawing}. The observer is rotating from the upper right to the upper left in this drawing and as the observer rotates the hot side of the warp comes directly into view. This increases the short wavelength flux, and since the orientation relative to the cold outer disk does not change with $\alpha$ the  long wavelength flux does not change significantly. The distance between the warp and the star keeps the warp cold and causes the change in the SED to be at longer wavelengths. Figure~\ref{warp_prec} also shows the precession of a warp seen at $i=85^{\circ}$. The infrared excess is smaller, because of the more highly inclined observing angle, while the change in flux is larger although the wavelength dependence is the same. As the disk becomes more inclined the warp is seen closer to face-on, creating more significant variations as it rotates in and out of view. 

Creating a warp in the middle of the disk is possible with a (sub)stellar companion embedded in the disk. If the companion is not coplanar with the disk then the gravitational perturbation from the companion can drive material out of the midplane. The disk around $\beta$ Pic displays a warp in the middle of the disk due to an unseen companion \citep{aug01}. The variable star WL 4 displays periodic variability in the near-infrared, indicative of a multiple system, as well as mid-infrared variability, indicative of a change in the dust structure near the star \citep{pla08}. Roughly 10\% of stellar systems have a binary companion within 1 AU \citep{mat94}. If the companion is within one AU of the central star then the precession period of such a configuration is years to decades, depending on the exact configuration of the system \citep{cap06}. The combination of the small flux change and long timescale would make precession difficult to observe.

We next consider the effect of changing the height of the warp. As the warp height increases the disk will get heated to a higher temperature since it is more directly illuminated by the radiation field. This is the same reason that a flared disk gets heated to a higher temperature and has a higher flux than a flat passive disk. The larger warp will also cast a larger shadow on the outer disk which will decrease its temperature and lower its flux. If the warp changes substantially enough then the change in the temperature structure would be evident in the SED. The hot surface layers of the disk that are seen in the mid-infrared respond almost instantaneously to a change in the stellar radiation field \citep{chi97}. This means that the timescale for the SED to change will depend on how quickly the disk can change its structure, which is the dynamical timescale. For the central star considered here the dynamical timescale can be as low as one week within 0.1 AU.

In order to examine this effect we take the disk we used to study precession ($R_{warp}=0.5$AU) and vary its scale height, parameterized by the factor $g$, from 0.01-0.1. This corresponds to a physical height of 0.005-0.05 AU. The results are shown in figure~\ref{warp_grow_i60}. We find that as the height of the warp increases the flux at $\lambda<15\micron$ increases while the flux longward of 15\micron\ decreases. The flux variations is small but the unique wavelength dependence would make infrared variability due to a changing scale height at the inner edge of the disk easy to diagnose. 

The same mechanism that drove the disk to become warped may also cause it to change the height of the warp. If the warp is caused by a companion, then this object could drag material as its orbit takes it out of the midplane \citep{fra09}. For a binary companion whose orbit is misaligned with the disk, the plane of the disk will equilibrate with the plane of orbit over thousands of years, but on the timescale of a single orbit dust will be shifted by the passing of the companion. As the companion's orbit takes it out of the plane of the disk it will drag material with it, but when the companion's orbit takes it back into the midplane of the disk this dust will quickly settle down. This would lead to periodic variations in the scale height of the warp on top of the long term variations. It may also be possible that a long-lived asymmetric structure within the disk, such as a warp caused by a companion, is experiencing variable illumination from the central source. This is possible if the accretion flow is localized into hot spots that rotate around the star. As the hot spot illuminates the warp, it will heat the side facing the star, causing its scale height to increase. The scale height of the warp will decrease as the hot spot rotates around to the far side of the star.

\subsection{Inner Warp}
As the warp moves closer to the star, the deviation in the SED from a flat passive disk moves to shorter wavelengths. To affect the flux at $5-8\micron$ we need to move the warp almost to the inner edge of the disk. We could also consider a warp that reaches its maximum height above the midplane at the inner, rather than the outer, edge of the disk. An inner warp has been invoked to explain the variability observed in AA Tau \citep{bou03} in which the stellar magnetic field is misaligned with the disk causing it to warp as material flows onto the field lines. A inner warp can be described by the function:

\begin{equation}
h(r,\theta)=gr_{min}\left(\frac{r}{r_{min}}\right)^{-4}\cos(\theta)
\end{equation}

\begin{figure}
\epsscale{.5}
\plotone{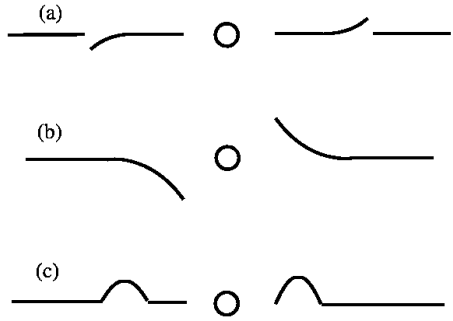}
\caption{Schematic drawing of our three simple geometric models. The three models are (a) the middle warp, (b) the inner warp and (c) the spiral wave all viewed edge on along $i=90^{\circ}$, $\alpha=-90^{\circ}$. \label{drawing}}
\end{figure}

\begin{figure}
\epsscale{1.}
\includegraphics[width=0.33\textwidth]{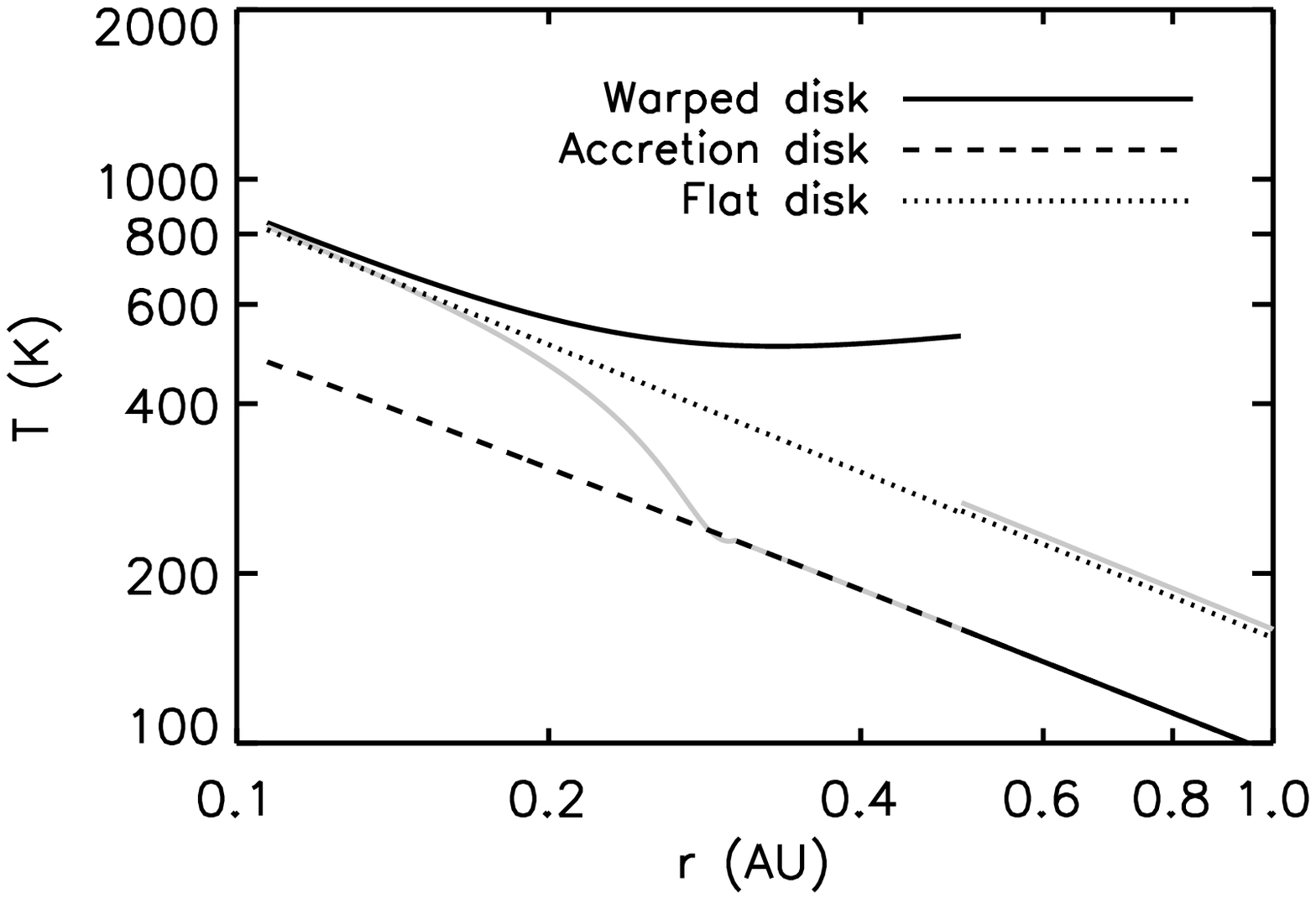}
\includegraphics[width=0.33\textwidth]{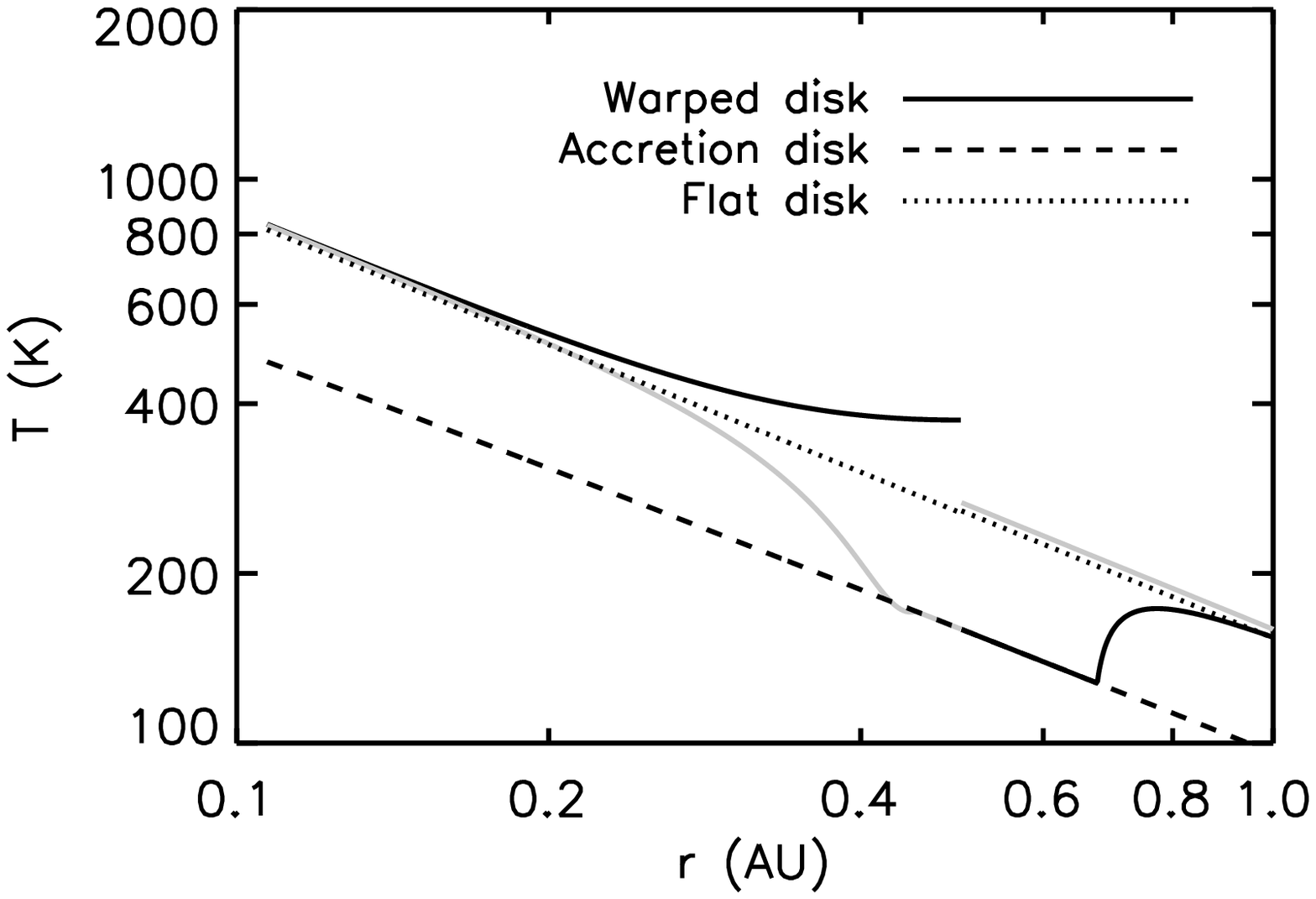}
\includegraphics[width=0.33\textwidth]{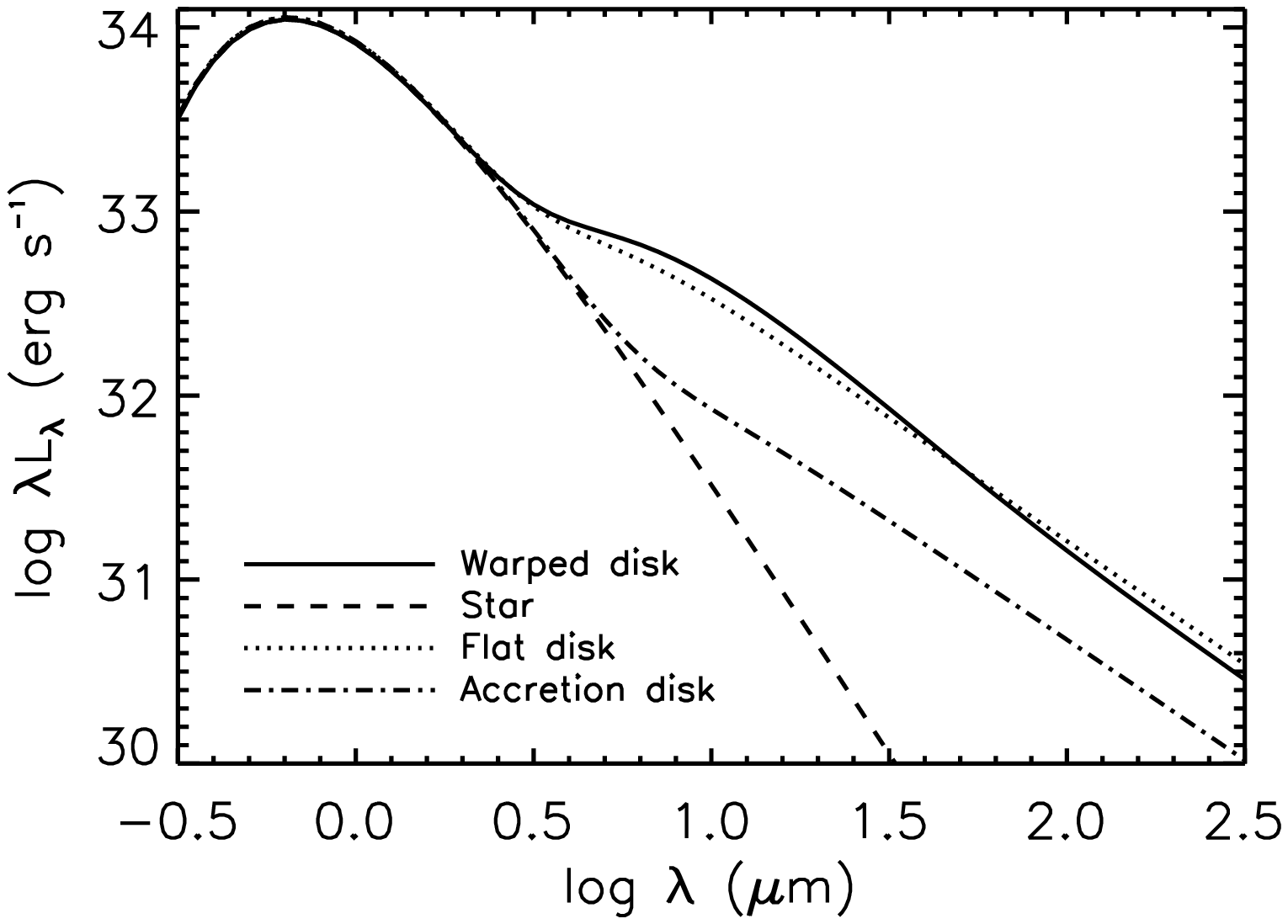}
\caption{Middle Warp disk. In the left and middle panels are the temperature structures of a disk with a warp that reaches its maximum h/r of $g=0.05,0.01$, respectively, at $r=0.5$AU with a flat disk extending beyond the warp to 100 AU. The dark solid line is the side of the disk facing the star, while the grey solid line is the side of the warp facing away from the star. On the right is the SED for a disk with an middle warp that grows to $g=0.05$ at $R_{warp}=0.5$ AU and a flat extension to the disk that extends to 100 AU. The disk is viewed along $i=60^{\circ},\alpha=0^{\circ}$. Also shown are a flat accretion disk, a flat passive disk and a blackbody at $T_{eff}=5700K$ for comparison. \label{warp1}}
\end{figure}

The temperature distribution for disks with $g=0.05,0.01$, $r_{min}=8R_*$ are shown in figure~\ref{warp2}. The exponent for the power law was chosen to match the middle warp and depends on the exact physical cause of the warp. The inner warp can be split into two pieces, a convex piece which directly faces the star and a concave piece which is turned away from the star. In the schematic drawing of the inner warp in figure~\ref{drawing}b on the right side the convex piece is on the bottom while the concave piece is on the top. The inner edge of the convex side reaches a very high temperature and departs significantly from a flat passive disk. The concave side just behind the warp has a low temperature because the warp completely blocks the star and the disk is only heated by viscous dissipation. Moving outwards, a point on the disk will be able to see some of the star, although it is still partially blocked. The temperature on the concave side far from the warp will be between that of a flat passive disk which experiences no shadowing and a fully shadowed disk which is only heated by viscous dissipation. As the warp grows the temperature on the convex side increases and on the concave side it decreases. In limit of a very large warp the concave side will be completely shadowed while the convex side will be much warmer than a flat passive disk.

Figure~\ref{warp2} compares the SED of the inner warp disk to that of a flat passive disk and a disk heated by viscous dissipation. The warped disk significantly departs from the flat passive disk at short wavelengths because of the high temperature on the convex side while its SED is lower at long wavelengths because of the shadowing on the concave side. The warp drops below the midplane for $90<\theta<270^{\circ}$ (the left side of the top half of the disk shown in figure~\ref{drawing}b) and only half of the disk is shadowed. Since at most half of the disk can be shadowed, the flux always lies above a flat accretion disk even for very large warps. 

Figure~\ref{warp2_prec} shows the variation in the SED as the warp moves around the star. The short wavelength flux changes as the convex side rotates more directly into the line of sight while the long-wavelength flux does not change because this part of the disk is essentially flat and the SED of a flat disk does not depend on azimuthal viewing angle.  If the warp does not occult the star (left panel of figure~\ref{warp2_prec}) there is a small change in the flux. Viewing the disk at a high enough inclination (right panel of figure~\ref{warp2_prec}) changes the form of the SED variations. The infrared variations are smaller when the disk is viewed edge on than when it is seen closer to face on. This is because when the warp on the near side of the disk blocks the star, the hot side of the warp on the far side of the star is directly visible. These two effects work in opposite direction; blocking stellar flux reduces infrared flux, while seeing the convex side of the warp increases the flux. This can cause the change in optical flux and infrared flux to be in opposite directions, while for the more face-on case the flux variations are in the same direction at all wavelengths. 

The type of warp described here can be created by a number of different physical mechanisms. One possibility is a (sub)stellar companion embedded in the disk. As with the middle warp, if the companion is not coplanar with the disk then the gravitational perturbation from the companion can drive material out of the midplane. If the companion is located close to the star then this disturbance will be at the inner edge of the disk. In the models of \citet{lar97}, for small mass ratios (q=0.01-0.1), the warp changes the scale height of the disk by $\Delta h/r\approx0.1$, similar to the scale height considered here. 
	
For the T Tauri star AA Tau, a warped disk caused by a tilted stellar magnetic field is believed to explain the observed variations \citep{bou03}. The stellar magnetic field is misaligned with the disk and the flow of material onto the magnetic field warps the disk. This warp would occur near the corotation radius where the magnetic field from the star truncates the disk, where the dynamical timescale of the disk will only be a few days. Interferometric observations generally find that the hottest circumstellar dust is at distances consistent with the dust sublimation radius, much larger than the corotation radius for higher mass stars\citep{mil07}. The exact location of the dust sublimation radius depends on the size of the dust grains and the geometry of the inner disk. If the disk is close to flat, as is assumed in our models, or consists of large grains, then it may be possible for the dust to extend close enough to the star to trace a warp caused by a tilted magnetic field.

The timescale for the inner warp to rotate around the star depends on the process originally responsible for creating the warp. If the motion is precession driven by a companion on a misaligned orbit then the period will be 1.5 years for a warp at 0.1 AU \citep{cap06}. A warp created by a tilted magnetic field will rotate at the same rate as the star.  Typical rotation periods for pre-main sequence stars are a few days to a week \citep[e.g.][]{coh04}, which would produce rapid flux variations. 

A variable stellar accretion rate could also change the structure of the inner disk. The height of the inner rim is set by the incident luminosity, which is a combination of stellar and accretion luminosity \citep{muz03}. Increasing the accretion luminosity will raise the temperature of the inner rim causing its scale height to increase. If the accretion luminosity is comparable to the stellar luminosity then a rapidly varying accretion luminosity could lead to a rapid change in the inner rim height. The accretion rate onto young stars varies rapidly on short timescales \citep{har91,gul96}. If the accretion luminosity was not evenly distributed around the star, but instead confined to a hot spot then this may lead to preferential heating and growth of part of the inner disk. Growing part of the inner disk creates a non-axisymmetric structure that may be comparable to the inner warp structure, producing similar changes in the SED. 

The growth of an inner warp can create an anti-correlation in the disk flux as it did with the middle warp. In figure~\ref{warp2_grow_i60} we show a model for $g=0.005-0.1 (h=0.005-0.01 AU)$ at $i=60^{\circ}$. As the warp grows the short wavelength flux increases while the long wavelength flux decreases due to the shadowing effect discussed above for the middle warp. The large temperature change in the inner disk leads to the significant variation in the near and mid-infrared flux. This flux variation would be easily observable in the near and mid-infrared.

The timescale to change the structure of the disk is the dynamical timescale, equal to the keplerian period, which is roughly one week at 0.1 AU. After the dust moves within the disk, its temperature will readjust to its new orientation relative to the radiation field on a thermal timescale. For the surface layers of the disk responsible for the mid-infrared emission this timescale is a few seconds, effectively instantaneous compared to the dynamical timescale \citep{chi97}. We compute warps of different sizes independently of each other, which is the same as assuming that as the warp grows its temperature responds instantaneously to the changing intercepted radiation field. These rapid variations in the scale height could be caused by the motion of a companion misaligned with the disk, as in the middle warp \citep{fra09}. 

\subsection{Spiral Wave}
We can also use the framework of TB96 to derive the SED of a simple spiral wave. Spiral waves are a common feature in the disks surrounding stars with planets or a stellar companion \citep{gol79,kley99}. The shocks created by the passing density waves can change the scale height of the disk \citep{edg08,bol06} causing heating and shadowing effects similar to a warp. We use a simple parameterization of a spiral wave as a modification of a flat disk to study its effects on the SED. The position of the spiral wave is given by:

\begin{equation}
r_{sw}=r_{min-sw}(1+n\theta)
\end{equation}

The value of $n$ sets the maximum radius of the wave and we keep $n$ fized at $0.8/(2\pi)$ to match more detailed descriptions of a spiral wave \citep{ogi02}. The parameter $r_{min-sw}$ sets the minimum radius for the center of the wave, which is taken to be slightly larger than the innermost radius of the disk. Our wave only runs from $0<\theta<360^{\circ}$, although it could continue to wrap around the star until it reaches the outer edge of the disk. The height above the midplane is given by:

\begin{equation}
h(r,\theta)=g\left(1-\frac{m\theta}{2\pi}\right)r_{min-sw}\exp\left(-\left(\frac{r-r_{sw}}{\sigma_r}\right)^2\right)
\end{equation}

The shape of the wave is a gaussian with width $\sigma_r$ centered at $r_{sw}$. The function $\left(1-\frac{m\theta}{2\pi}\right)$ modifies the height of the wave so that it decreases as it extends around the disk. Spiral waves grow weaker as they get further from their point of origin and the parameter $m$ can be used to modify how quickly the wave height decreases. In our fiducial model the wave disappears at $2\pi$ ($m$=1). The factor $g$ serves a similar purpose as before; it sets $h/r$ for the wave. Although this parameterization allows us to modify many details of the spiral wave we generally keep $m=1, \sigma_r=0.1r_{min}, r_{min-sw}=r_{min}+2*\sigma_r,n=0.8/2\pi$ and modify $g$.

The temperature structure of spiral disks with $g=0.05,0.1$ is shown in Figure~\ref{warp3}. On the side of the wave facing the star the disk reaches a very high temperature, while directly behind the wave the temperature drops. Further out from the wave the disk is only partially shadowed and the temperature lies between that of a flat passive disk and the base flat accretion disk. In the models shown in figure~\ref{warp3} the radius of the wave increases with $\theta$ while the height decreases, hence at $\theta=180^{\circ}$, shown as the grey line, the temperature is lower on the front side of the wave and higher behind the wave than for $\theta=0^{\circ}$ . Comparing the two models, we find that as the size of the wave increases the temperature of the side of the disk facing the star increases while the outer disk becomes more shadowed.

The resulting SED for $g=0.05$ is shown in Figure~\ref{warp3}. It rises above the flat passive disk shortward of $\log\lambda=1.0$ while longward it drops significantly below the flat passive disk. The short wavelength increase over a flat passive disk is smaller for the spiral disk than for the inner warp because less of the disk is heated up to a high temperature (compare Figure~\ref{warp2} and ~\ref{warp3}). The decrease at long wavelengths is much larger because in the outer disk every $\theta$ is exposed to a wave inside its radius. For the warp only half of the disk is shadowed at one time, while the spiral wave can shadow all of the disk (figure~\ref{drawing}). Comparing the temperature of the outer convex side of the inner warp disk with the $\theta=180^{\circ}$ spiral wave we can see that the temperature is lower for the spiral wave, which will result in a lower flux at longer wavelengths.

An embedded massive planet within the disk will create a spiral wave \citep{pap07}. These waves drive shocks which can substantially increase the scale height of the disk \citep{bol06,edg08}, up to $\Delta h/r\approx0.6$ \citep{bol06}. The shadow cast by the planet on the outer disk could be imaged in scattered light, but would be difficult to observe in the infrared \citep{jan09}. The presence of massive planets have often been invoked to explain the lack of short-wavelength excess within young circumstellar disks, and spiral waves are a natural extension of this situation. A spiral wave created by a planet embedded in the disk will stay fixed relative to the planet and will travel around the star at the orbital frequency of the planet. For a planet within 0.1 AU of a star this can approach one week, which is much shorter than the precession of the warped disks. We can demonstrate the change in SED as the planet orbits around the star by observing the spiral wave at different azimuthal angles, shown in figure~\ref{warp3_prec}. The flux changes by $<0.05$ dex and the change in the short and long wavelength flux is correlated, when the disk is viewed close to edge on. The small change in flux is due to the spiral wave being closer to an axisymmetric structure than either the middle warp or the inner warp. If the disk is highly inclined then the spiral wave will occult the star and this occultation will dominate the variations in the infrared (right panel of figure~\ref{warp3_prec}). Here the infrared variations are substantial and strongly dependent on viewing angle, but are mainly due to changing the underlying flux from the star rather than changing the flux from the disk. 

Varying the wave height will change the shadowing of the disk and can produce an anti-correlation in flux (Figure~\ref{warp3_grow_i60}), as in the previous models. In these models ($g=0.005-0.1$, $h=0.0006-0.013$ AU) we view the disk along $\alpha=180^{\circ}$ in order to directly observe the warmest part of the wave and produce the largest mid-infrared variations. As the wave grows the short-wavelength ($\lambda<10\micron$) flux does not change as much as the inner warp because only a small part of the disk will be heated as it grows. The extra heated part of the spiral disk is confined to the inner edge of the wave, whose position does not change. When the inner warp grows, more of the disk is directly exposed to the star. This means that not only does the temperature at one point rise, but the area of the disk that departs from a flat disk is larger. The spiral disk only has the effect of increasing the temperature at a point, without increasing the area over which the disk is warmer than a flat disk. Beyond 10\micron\ the disk flux drops dramatically with an increase in the height of the wave. This is because all angles $\theta$ are subject to being shadowed; for the inner warp only $-90^{\circ}<\theta<90^{\circ}$ will be shadowed. This changes can reach up to $\sim0.2$dex at 30\micron. Changes in the height of the wave may be caused by the nonlinear shock as it passes through the disk \citep{bol06} or it may be caused by localized accretion hot spots heating different portions of the wave, as discussed for the warped disk models.

Our spiral wave model can also be used to simulate other disturbances in a disk that appear as a wave. One such situation is a thermal instability in which a vertical disturbance grows in to a wave that slowly travels inward toward the star \citep{wat08}. In accretion disks the temperature and structure of the disk are closely connected since the scale height depends on the disk temperature at a given radius. If the disk is flared, it is more directly illuminated by the stellar radiation field and hence is warmer at a given radius than if it were flat. If a small disturbance raised a piece of the disk above the midplane it would heat up as it became more directly illuminated. The frontside of this disturbance will grow because it is warmer, while behind the disturbance the disk is shadowed and it will shrink. As a result, the disturbance will propagate towards the star as a wave. \citet{wat08} find that this type of instability can develop in the outer disk and is long lived enough to travel within a few AU of the star.

We treat the instability as a spiral wave whose radial position and height do not change as a function of $\theta$ within the disk ($n=0$, $m=0$), and compare three locations and heights of the instability (r=0.5,0.4,0.3 AU with g=0.05,0.02,0.01 respectively) that are typical of thermal instability models \citep{dal99}. The results, shown in figure~\ref{l31_warp3_ti} have a pivot point in the SED at much longer wavelengths than the other models. At $>$3\micron\ the flux change is $\sim$0.1 dex until the far-infrared where the flux change can be much larger. In the models with $g=0.05,0.02$ the entire outer disk is shadowed and there is very little change in the far-infrared flux between these models. The flux from the outer disk only changes for the g=0.01 model where less of the disk is in shadow, which increases its flux. These waves travel inward on a thermal timescale, which is on the order of years.

\section{Comparison to Observations}
\subsection{LRLL 31}
 \citet{muz09} suggested that the mid-infrared variability of LRLL 31 may be due to a non-axisymmetric structure in the inner disk, and with our simple models we can more directly test this hypothesis. The observations of LRLL 31, described in detail by \citet{muz09} and summarized here, include multiple epochs with the {\it Spitzer} IRAC \citep{faz04}, MIPS \citep{rie04} and IRS \citep{hou04} instruments. Originally identified as variable based on two earlier IRAC and MIPS surveys of the field, it was followed up with further imaging and spectroscopic observations in the fall of 2007 and the spring of 2008. The entire set of data are shown in SED form in Figure~\ref{l31obs}. IRS spectra taken a week apart (October 9,16 2007) show that the 5-8\micron\ flux decreases by 0.2 dex, while the 12-40\micron\ flux increases by a similar amount. After four months (February 24, 2008) the 5-8\micron\ flux has risen, with a corresponding decrease in 12-40\micron\ flux, close to its original state. Another IRS spectra taken a week later (March 2, 2008) show the disk still in its high 5-8\micron\ state with the same anticorrelated change in the short and long-wavelength flux. IRAC observations taken from the GTO \citep{lad06} and C2D \citep{jor06} surveys (February 11, 2004 and September 8, 2004 respectively) also show substantial changes consistent with the IRS spectra. The MIPS 24\micron\ measurements include observations from the original GTO and C2D \citep{reb07} maps (February 21, 2004 and September 19, 2004 respectively), as well as more intensive observations from our GO program (PID 40372). The MIPS followup includes five consecutive days (September, 23-27, 2007) and two further maps the following spring (March 12,19, 2008). The 24\micron\ fluxes change by a few tenths of a magnitude on weekly to monthly timescales. The uncertainty in the infrared observations is $<5\%$ while the changes range from 20-60\% suggesting that the variations being observed are real. In Figure~\ref{l31obs} the October 9, 2007 spectra is shown with error bars to demonstrate the typical uncertainty in the spectra, which is smaller than the observed flux changes. A more detailed description of the data reduction, as well as the uncertainties in the data, is presented in \citet{muz09}. All of our observations show variations at all timescales over the entire observed wavelength range.
 
 The low 5-8\micron\ flux seen in the October 16, 2007 observation is difficult to explain with the standard axisymmetric puffed inner rim model. In that case, the amount of dust emission is set by the scale height, which in turn depends on the stellar plus accretion luminosity. In the limit of low $L_{acc}$ and/or high $L_*$, as is the case for LRLL 31, the irradiation of the inner rim cannot change significantly, and its emission should always be considerably in excess of the photospheric level. Disk models with large scale asymmetries such as warps or spiral waves, coupled with highly inclined viewing angles, offer a way around the problem.

We take LRLL 31 to be a G6 star, based on NIR spectra \citep{muz09}, with a radius of $2.9R_{\odot}$ and a mass of $1.8M_{\odot}$ derived using the \citet{sie00} isochrones. It has high polarization and its extinction is roughly $A_V=10$. When presenting the observations below, all of the data have been dereddened by $A_V=10$. \citet{dah08} find the accretion rate in the range $10^{-8}-10^{-7}M_{\odot}yr^{-1}$ based on optical veiling and various emission line diagnostics. The inclination of the disk has not been directly measured, but given the high extinction and high polarization, which are signs that some of the light passes through the disk, we focus on inclinations close to edge on.

Given the modeling results presented earlier, we focus on changing the height of the warp/wave. The middle warp, inner warp and spiral wave all produced a wavelength dependence similar to LRLL 31, where the short-wavelength flux would increase while the long-wavelength flux would decrease due to the shadowing of the outer disk. Figures~\ref{l31_warp_grow},~\ref{l31_warp2_grow} and ~\ref{l31_warp3_grow} show the middle warp, inner warp and spiral wave close to edge on with varying heights along with the observations of LRLL 31. An inclination of i=$85^{\circ}$ was chosen for these three models, while different values of $\alpha$ were chosen to maximize the flux change. The middle warp displays a similar wavelength dependence as the observations, but is unable to match the size of the variations at 3.6,4.5\micron. Also the location of the pivot in the SEDs is at a longer wavelength than observed.  The inner warp does a better job of fitting the size of the variation as well as the slope of the IRAC and short-wavelength IRS data. For the inner warp we consider two inclinations that are both close to edge-on but produce different SEDs. The difference in the two edge-on inclinations comes in the optical where the i=$85^{\circ}$ disk occults the star while the i=$95^{\circ}$ disk does not. This model also shows a pivot point in the SED, but it is also at a longer wavelength than observed. For the spiral wave model the pivot point moves to a wavelength that is close to 8\micron\ but the short wavelength flux variation is too small to match the observations, in particular the 3.6,4.5\micron\ photometry and the spectra with the lowest 5-8\micron\ flux. The timescale for the warp to change height, which is the dynamical timescale, approaches one week at 0.1 AU around LRLL 31. This is short enough to match the rapid variations seen in the mid-infrared. We do not consider the thermal instability because the timescale is much longer than one week.

Based on our simple models, we find that a disk with an inner warp whose height rapidly varies is able to best explain the observations of LRLL 31. It can fit the flux change at $\lambda<8\micron$, it includes a pivot point in the SED, and the timescale is consistent with the observations, but the simplicity of the model prevents it from fitting in detail.  More detailed radiative transfer models would help to further constrain the shape of the perturbation. Future work should also tie these parametric models to specific physical mechanisms in order to provide predictions for the detailed time-dependent behavior of the flux changes.

\subsection{Application to Other Variables}
In the previous models we considered a star whose mass was on the high end for a typical T Tauri star and a lower mass star with a lower effective temperature will not heat the disk to as high of a temperature at the same distance from the star, which will result in a weaker excess for a given value of h/r. As the disk precesses, only the orientation of the disk relative to the observer changes, which does not depend on the central star or the temperature at a particular location within the disk. The change in infrared excess with precession will then be similar over a wide range of masses. The same is true if we vary the scale height of the disk. In figure~\ref{tts} we show an inner warp seen at $i=60^{\circ}$ around LRLL 31, a typical T Tauri star (T$_{eff}$=4000 K, $M_*$=1$M_{\odot}$, $R_*=2R_{\odot}$,$\dot{M}=5*10^{-9}M_{\odot}yr^{-1}$, $R_{min}=8R_*$) and a brown dwarf (T$_{eff}$=3000 K, $M_*$=0.08$M_{\odot}$, $R_*$=0.7$R_{\odot}$, $\dot{M}=10^{-9}M_{\odot}yr^{-1}$, $R_{min}=8R_*$) with a height varying from $g=0.01-0.1$. The accretion rates were chosen to decrease with mass, but they only contribute significantly to the heating of the outer disk where our assumption of a flat disk severely underestimates the flux from typical disks. The change in the mid-infrared flux around all three stars is almost identical. While the exact structure and temperature of these three disks differs, as the mass decreases the temperature of the disk decreases and the physical distance between the disk and the star is smaller, the change in the warp is the same resulting in similar variations in the infrared flux. Changing from an inner warp to a middle warp or a spiral wave will also show a similar lack of dependence on the central source as the height of the warp/wave varies. If the structural change in the inner disk does not depend on the mass of the central star then such a change in the disk structure would be easily detectable in the near and mid infrared flux from many young stellar objects.

Since changing the spectral type of the central star does not have a significant effect on the infrared variability presented here, these models can provide insight into other variable objects.  \citet{mor09} use IRAC to survey 3-8\micron\ variability among young stellar objects within IC 1396A. They find that variability occurs in more than half of the young stellar objects, typically showing colorless variations with amplitudes of 0.05-0.2 magnitudes on timescales of a few days to a week. This is consistent with our models of a growing inner warp, which when the warp does not occult the star, can produce large variations in flux without a large change in the color in this wavelength range. \citet{hut94} find two stars whose BVRI fluxes and 4.5,10\micron\ fluxes vary in opposite directions while the near-infrared fluxes stay constant.  \citet{juh07} find similar results for SV Cep where the V band and 3.6\micron\ flux are anti-correlated. They also find that the V band and 100\micron\ flux are correlated with each other, while there is very little change in the flux at 25\micron. This is qualitatively similar to our model of a growing inner warp seen at$i=85^{\circ}$ (Figure~\ref{l31_warp2_grow}). As the warp grows the star becomes blocked, dropping the optical flux, and more of the outer disk is shadowed, dropping the long wavelength flux, producing a correlation between the optical and 100\micron\ flux. Also the larger warp emits more flux in the mid-infrared, producing an anti-correlation between the optical and 3-10\micron\ flux, while the near-infrared bands stay roughly constant. 

UX Ori stars are a type of optical variable that exhibit occasional dimming and reddening of their optical colors due to occultations of the star by clumps of gas and dust. These clumps are thought to arise from the inner edge of the circumstellar disk, which can occult the star since the system is seen very close to edge on \citep{dul03}. In some models of the infrared excess from these stars it has been assumed that the disturbance at the inner edge does not cover enough area to substantially shadow the disk \citep{pon07}. Our models show that there can be significant mid-infrared variations even if the perturbations at the inner edge of the disk do not shadow the entire disk.

Ongoing studies of mid-infrared variability will help to determine how common this type of variability is among young stellar objects. Multiple programs in progress gathering multi-epoch Spitzer data at 3.6 and 4.5\micron\ will constrain the frequency of mid-infrared variability among T Tauri stars. Given the wavelength coverage of these variability studies, and the expected mid-infrared variability due to non-axisymmetric variations in the inner disk, any changes in the structure of the inner disk will be readily observable. Using these observations  to look for periodicity, which is a sign of a perturbing companion, or a correlation with accretion rate will help to constrain the physical  processes that influence the inner structure of circumstellar disks.

\section{Conclusion}

We have used simple geometric models to investigate the effects of non-axisymmetric structure in a circumstellar disk on the system's SED. By simplifying the models we have been able to investigate a wide range of geometries including a warp in the middle of the disk, a warp at the inner edge of the disk and a spiral wave. These models show an anticorrelation between the flux at short and long wavelengths because of variable shadowing of the outer disk, in qualitative agreement with many observations. However, precession or corotation of these structures produces a negligible change in the infrared excess amplitude. Only by changing the scale height of the perturbations, as may be produced by the dynamical interactions or variable illumination of disk structures, do we find significant flux variations as observed. Further observations spanning a range of timescales and covering multiple objects will help to discriminate between models as well as indicate the frequency of such structural variations. The similarity in the size of the infrared variations across a wide range of masses also indicates that the type of non-axisymmetric variability explored here would be detectable around many young stellar objects. Numerous observations of mid-infrared variability around young stellar objects are consistent with our models.

\acknowledgments
We would like to thank George Rieke for helpful comments on the modeling and the anonymous referee for suggestions that substantially improved the organization and flow of the paper. This work is based in part on observations made with the Spitzer Space Telescope, which is operated by the Jet Propulsion Laboratory, California Institute of Technology under a contract with NASA. This work is supported in part by contract number 960785 issued by JPL/Caltech.

\begin{figure}
\plottwo{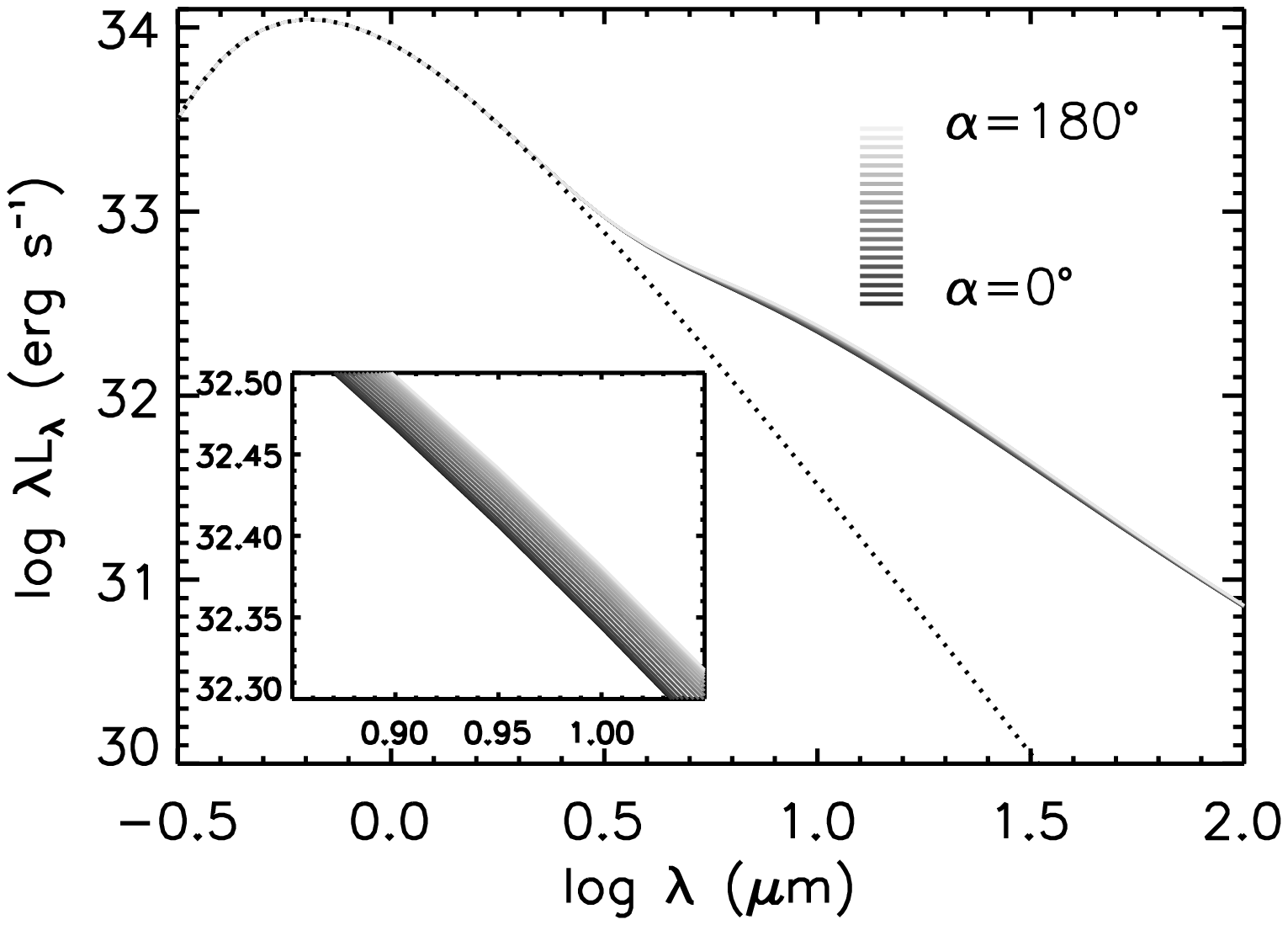}{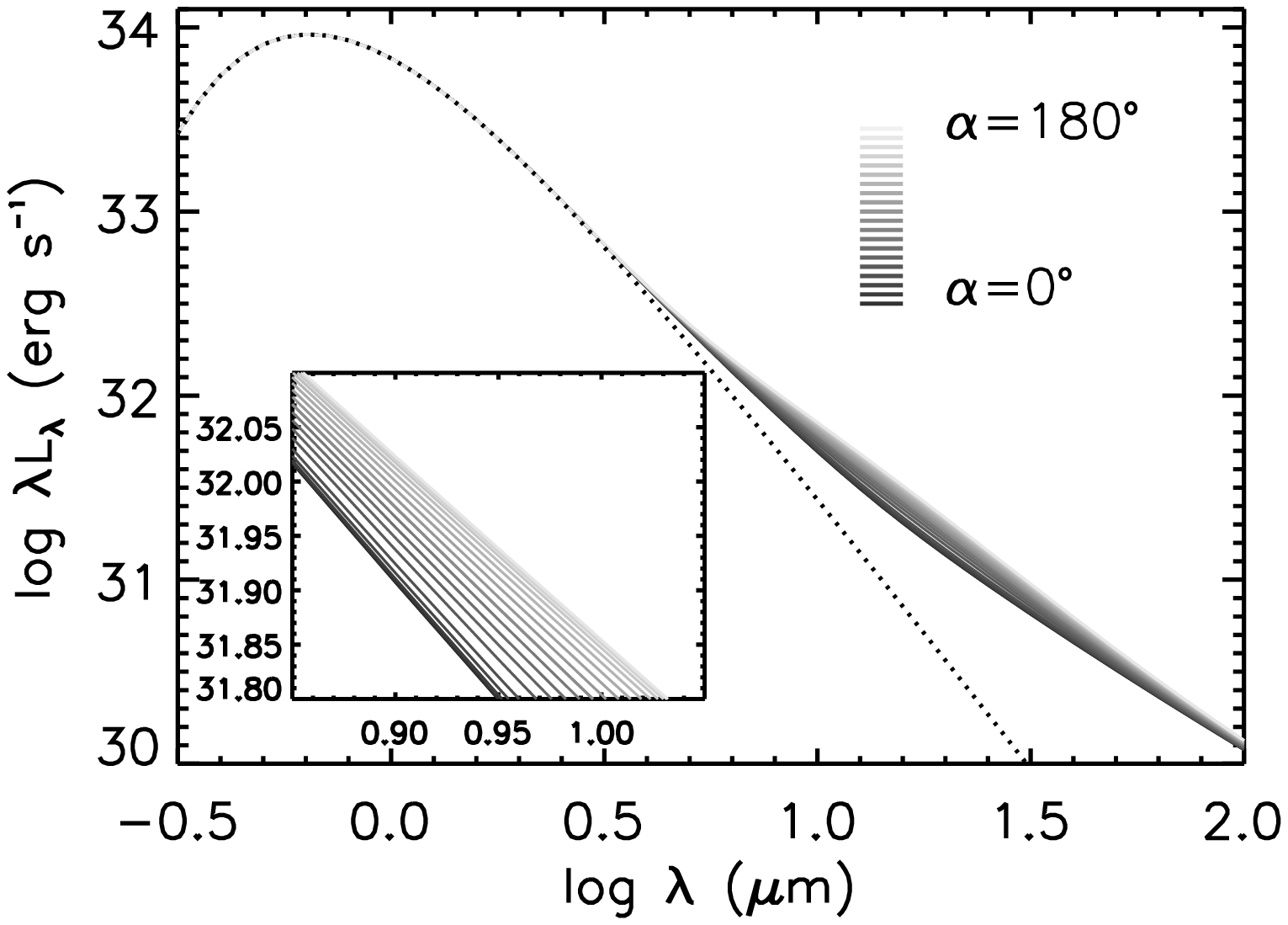}
\caption{SED of a precessing disk with an middle warp extending to $R_{warp}=0.5$ AU with $g=0.05$ and a flat extension seen at an inclination of $60^{\circ}$ (left panel) and $85^{\circ}$ (right panel). From dark to light lines the disk rotates through $180^{\circ}$. The inset shows a closeup of the models near 10\micron. The dotted line is the photosphere. The change in the infrared flux is small, but it increases as the disk is viewed closer to edge on. \label{warp_prec}}
\end{figure}

\begin{figure}
\plotone{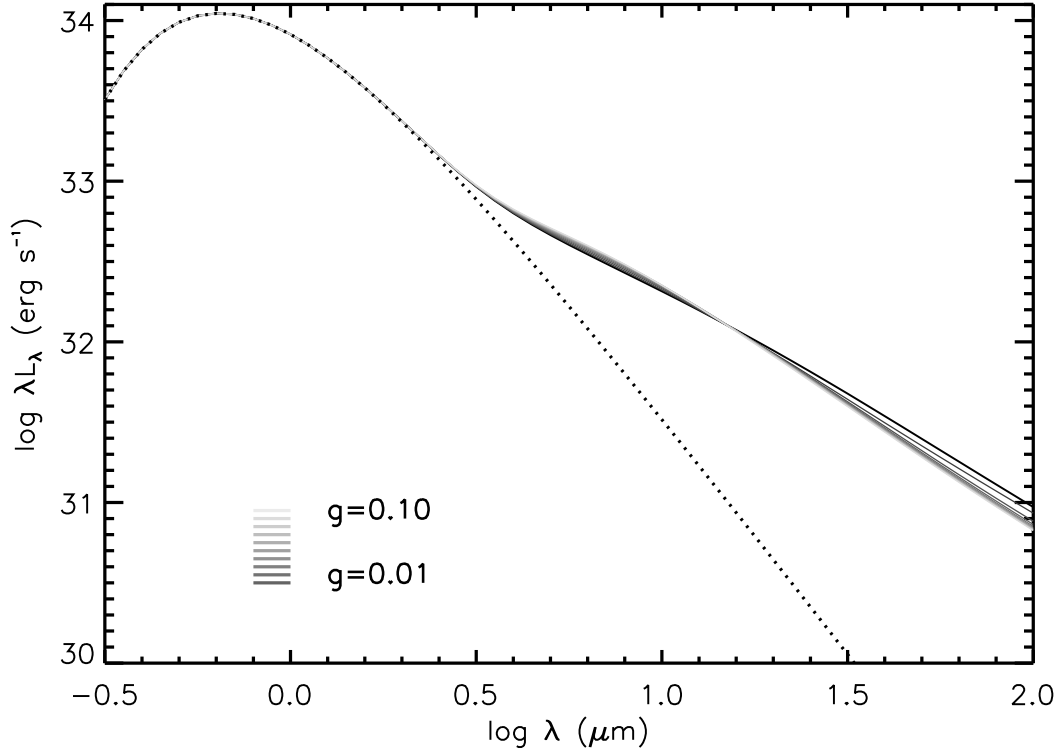}
\caption{SED for a growing warp ($g=0.01-0.1$ from dark to light lines) viewed at $i=60^{\circ},\alpha=0^{\circ}$.\label{warp_grow_i60}}
\end{figure}

\begin{figure}
\includegraphics[width=0.33\textwidth]{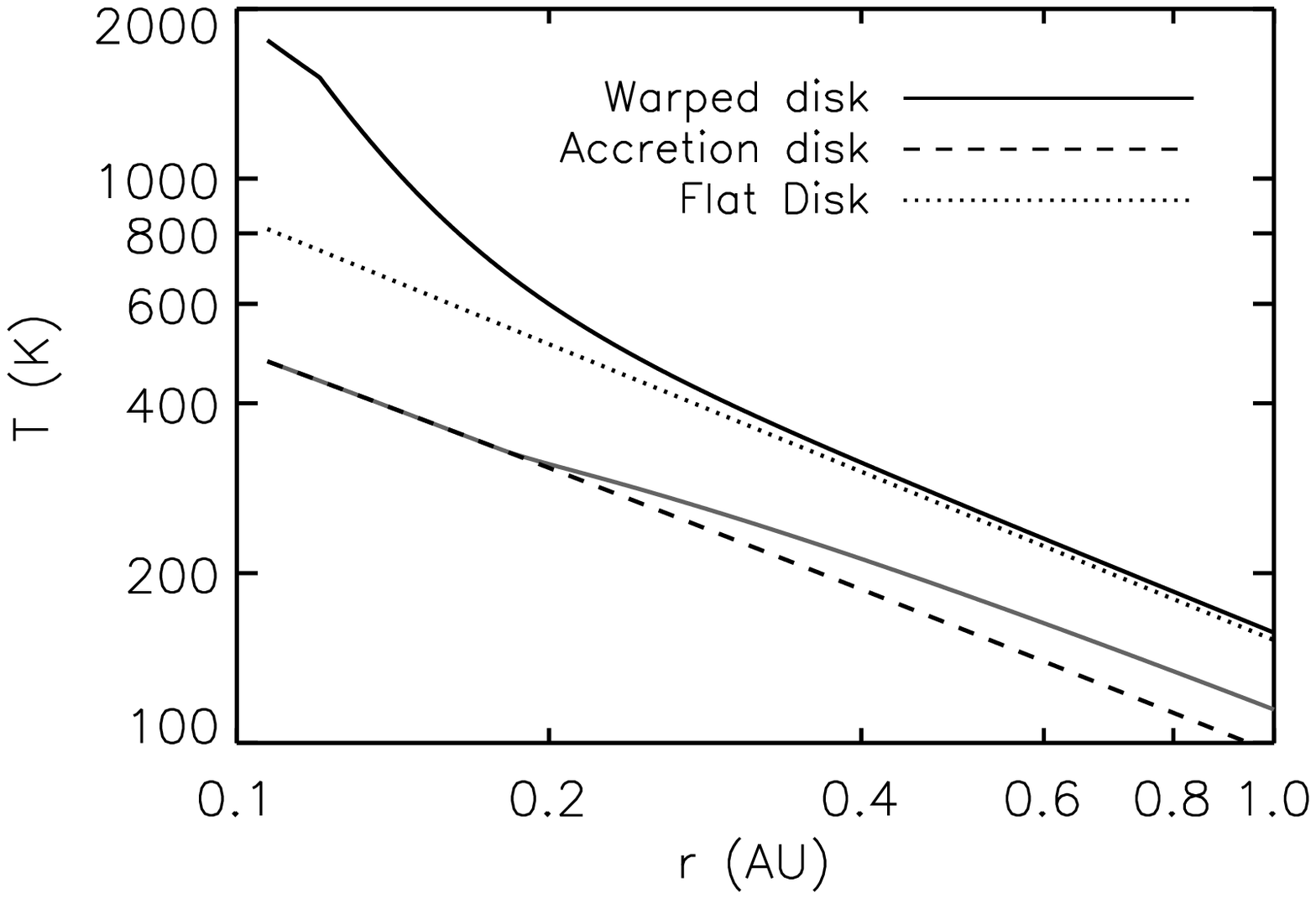}
\includegraphics[width=0.33\textwidth]{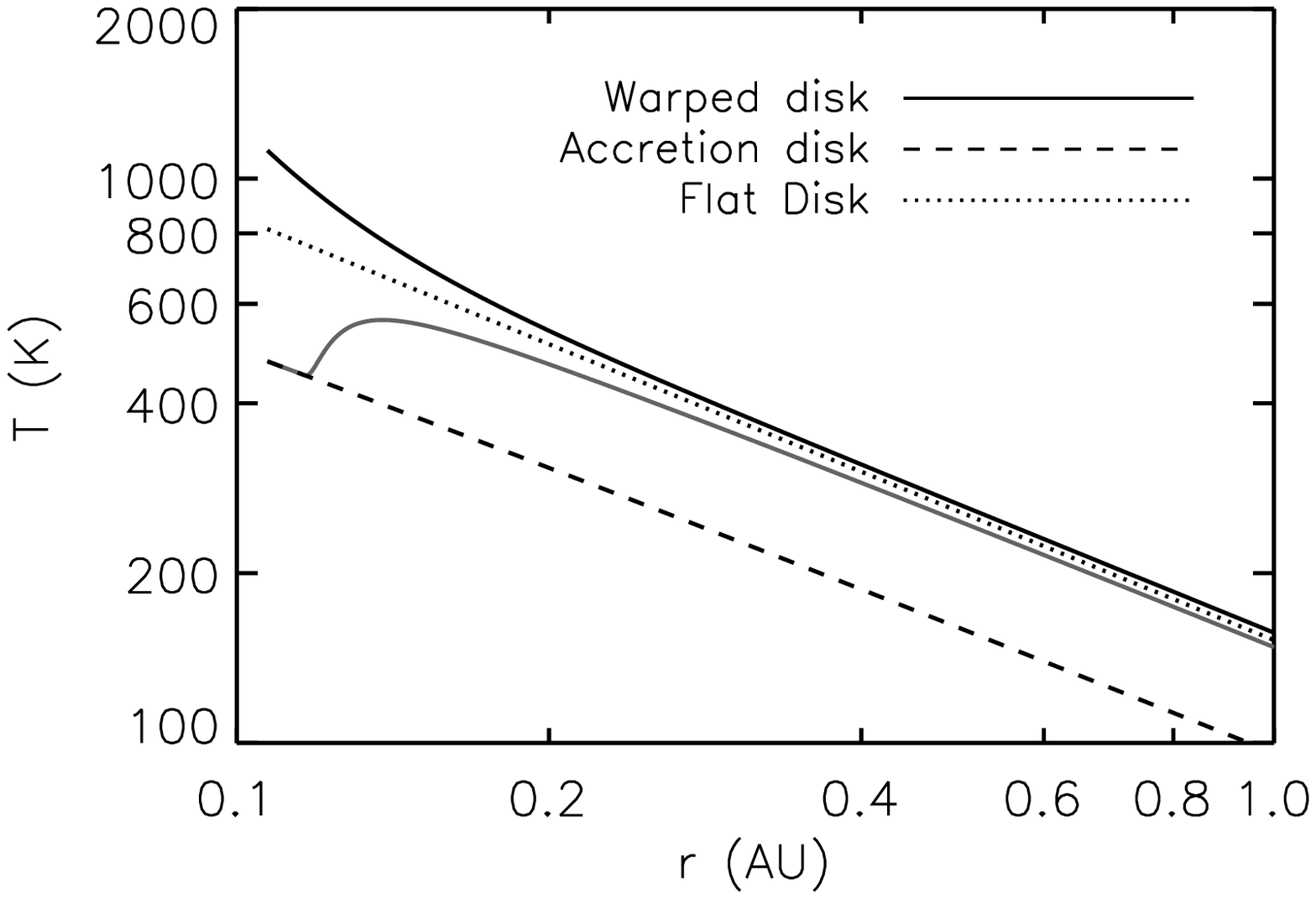}
\includegraphics[width=0.33\textwidth]{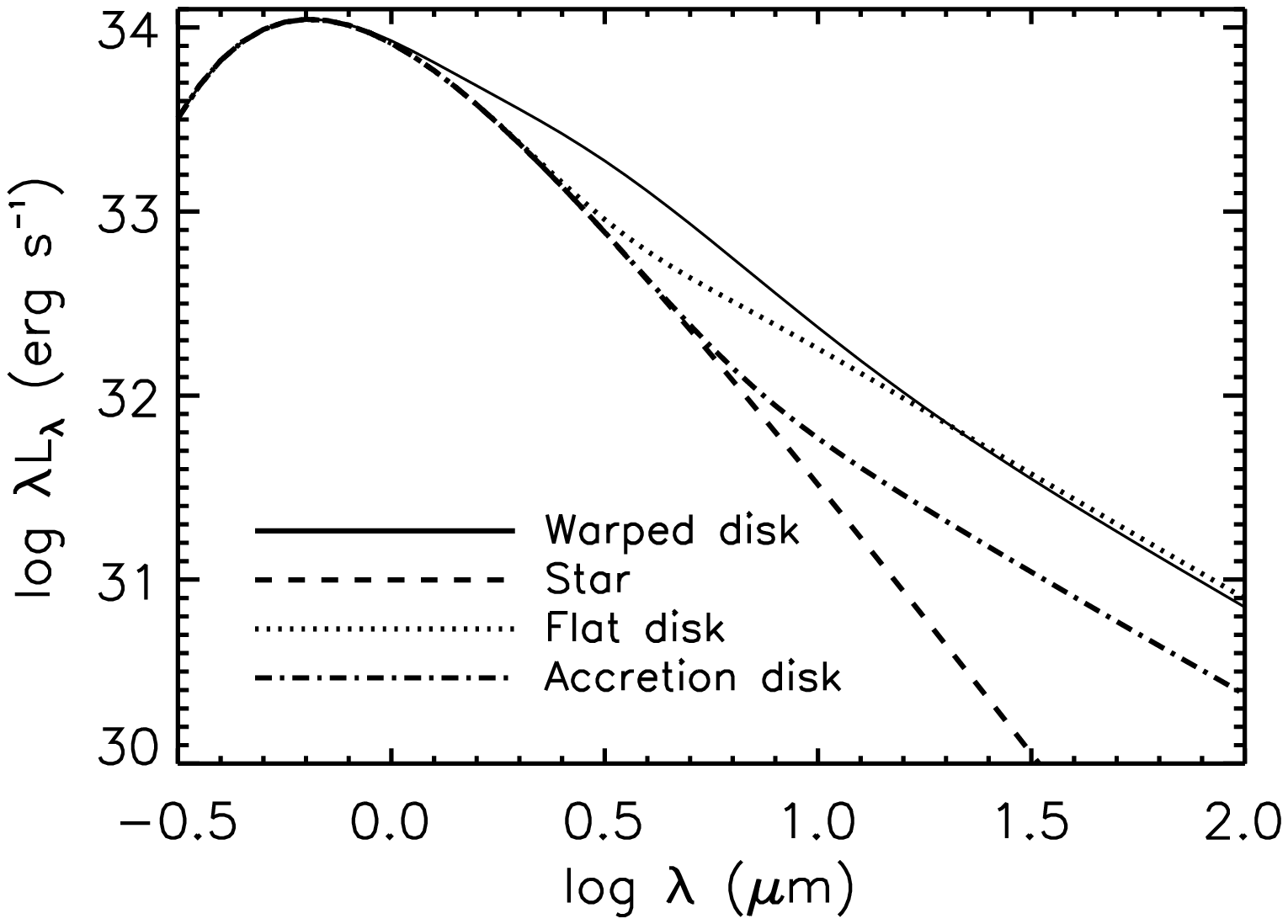}
\caption{Inner warp disk.  The left and middle panels show the temperature distribution for a disk with an inner warp of $g=0.05$ and $g=0.01$ respectively. The dashed line is the temperature of a disk heated solely by viscous dissipation ($\dot{M}=10^{-8}M_{\odot}yr^{-1}$), the dotted line is a perfectly flat passive disk, while the solid (dark and light) lines are the warped disk (convex and concave sides respectively). The maximum temperature of the warp, and the shadowing of the outer disk changes with the size of the warp. On the right is the SED for a warped inner disk with $g=0.05$ seen at $i=60^{\circ},\alpha=0^{\circ}$. For comparison a flat passive disk, a disk heated by viscous accretion ($\dot{M}=10^{-8}M_{\odot}yr^{-1}$) and the photosphere are also included.  \label{warp2}}
\end{figure}

\begin{figure}
\plottwo{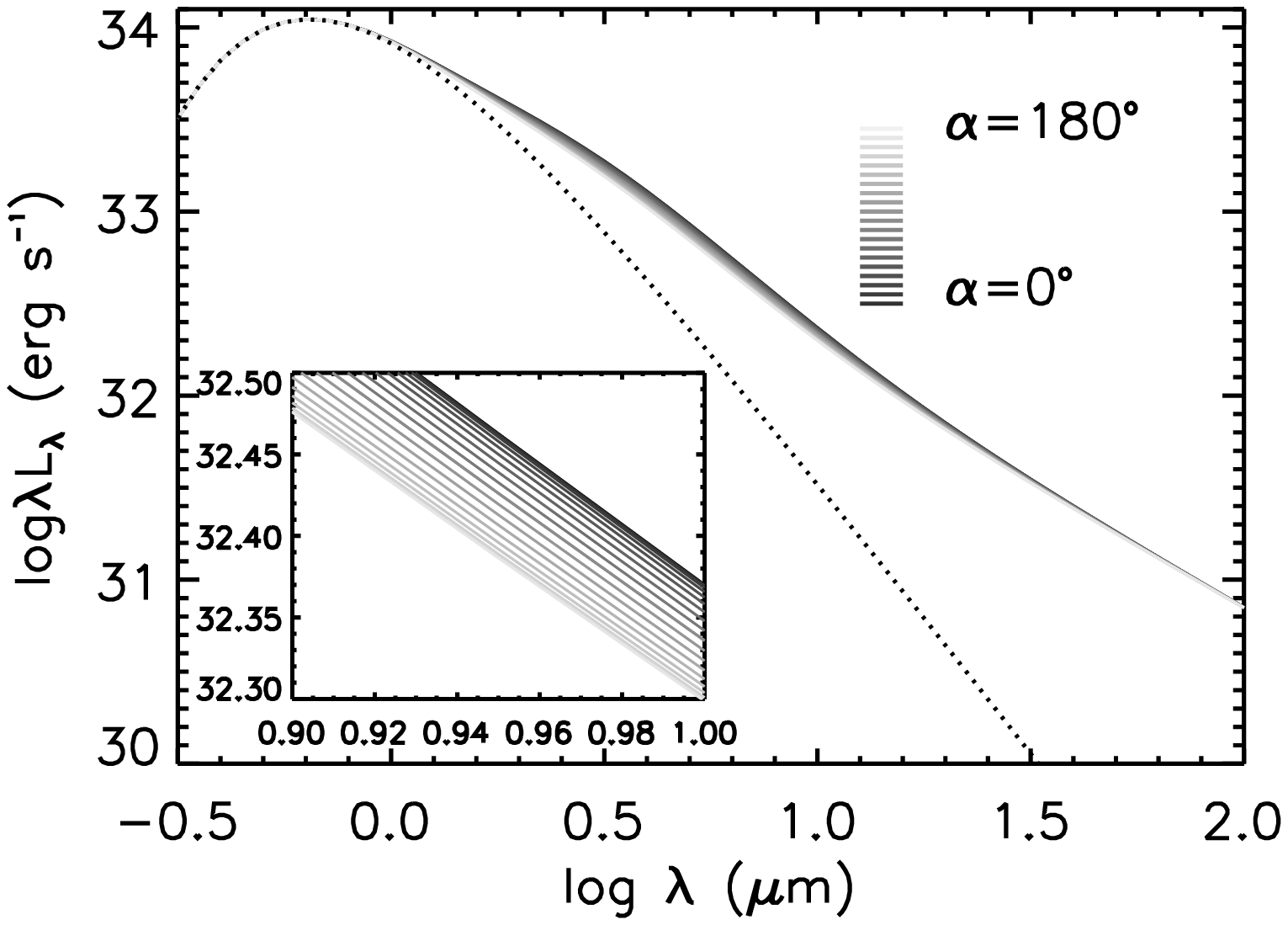}{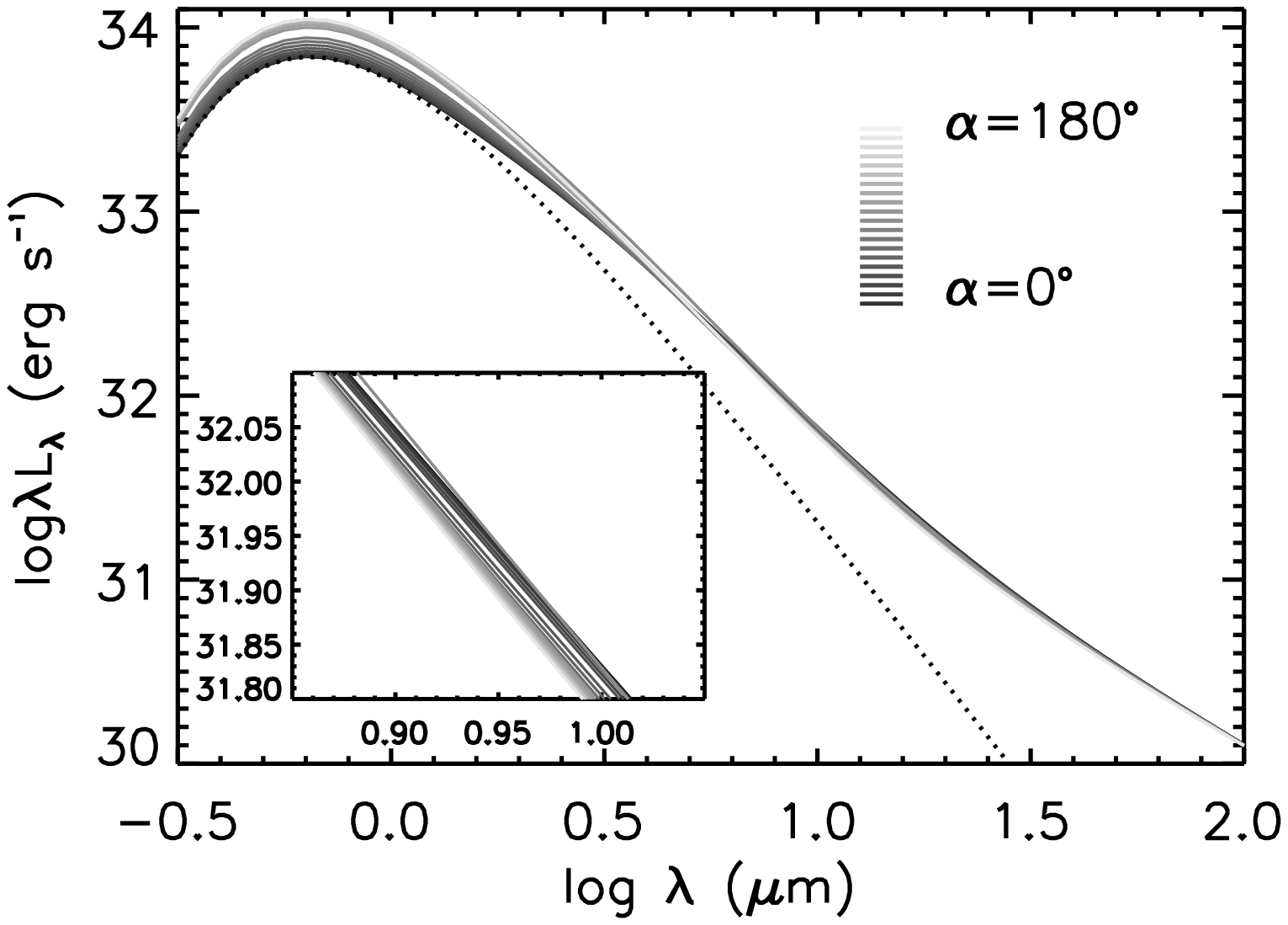}
\caption{SEDs for a warped inner disk with $g=0.05$ showing the change in flux as the disk precesses. From dark to light the disk changes orientation from $\alpha=0^{\circ}$ to $\alpha=180^{\circ}$ with an inclination of $i=60,85^{\circ}$ (left and right panels respectively).  \label{warp2_prec}}
\end{figure}

\begin{figure}
\plotone{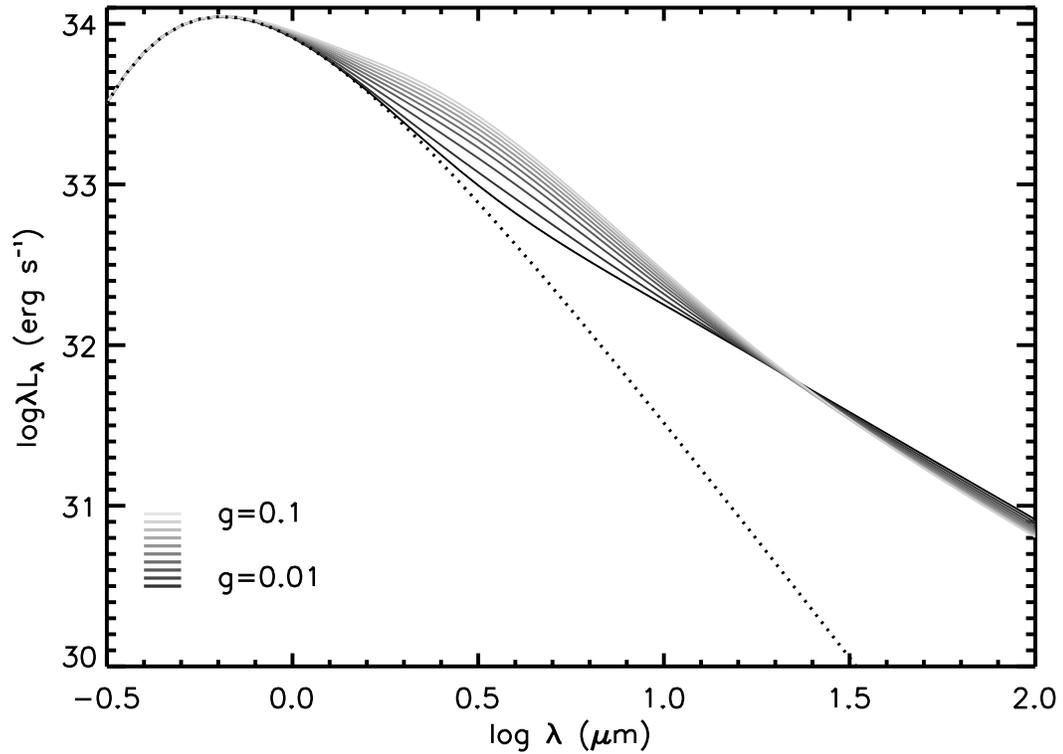}
\caption{SEDs showing a growing inner warp. The models vary $g$ from 0.01 to 0.1 (from dark to light lines) seen at an inclination of $i=60^{\circ}$. There is a substantial change in the mid-infrared flux as the warp grows \label{warp2_grow_i60}}
\end{figure}

\begin{figure}
\includegraphics[width=0.33\textwidth]{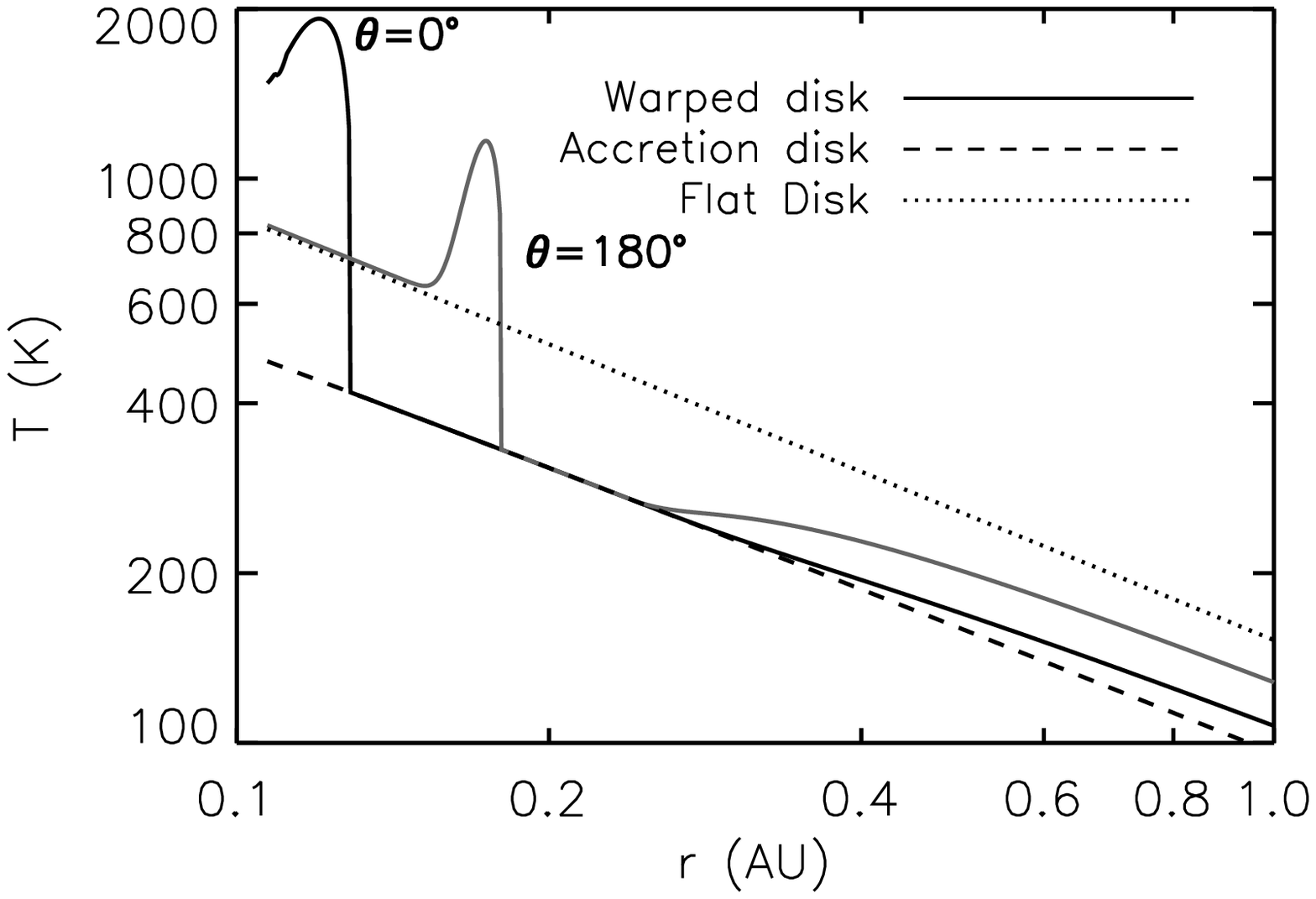}
\includegraphics[width=0.33\textwidth]{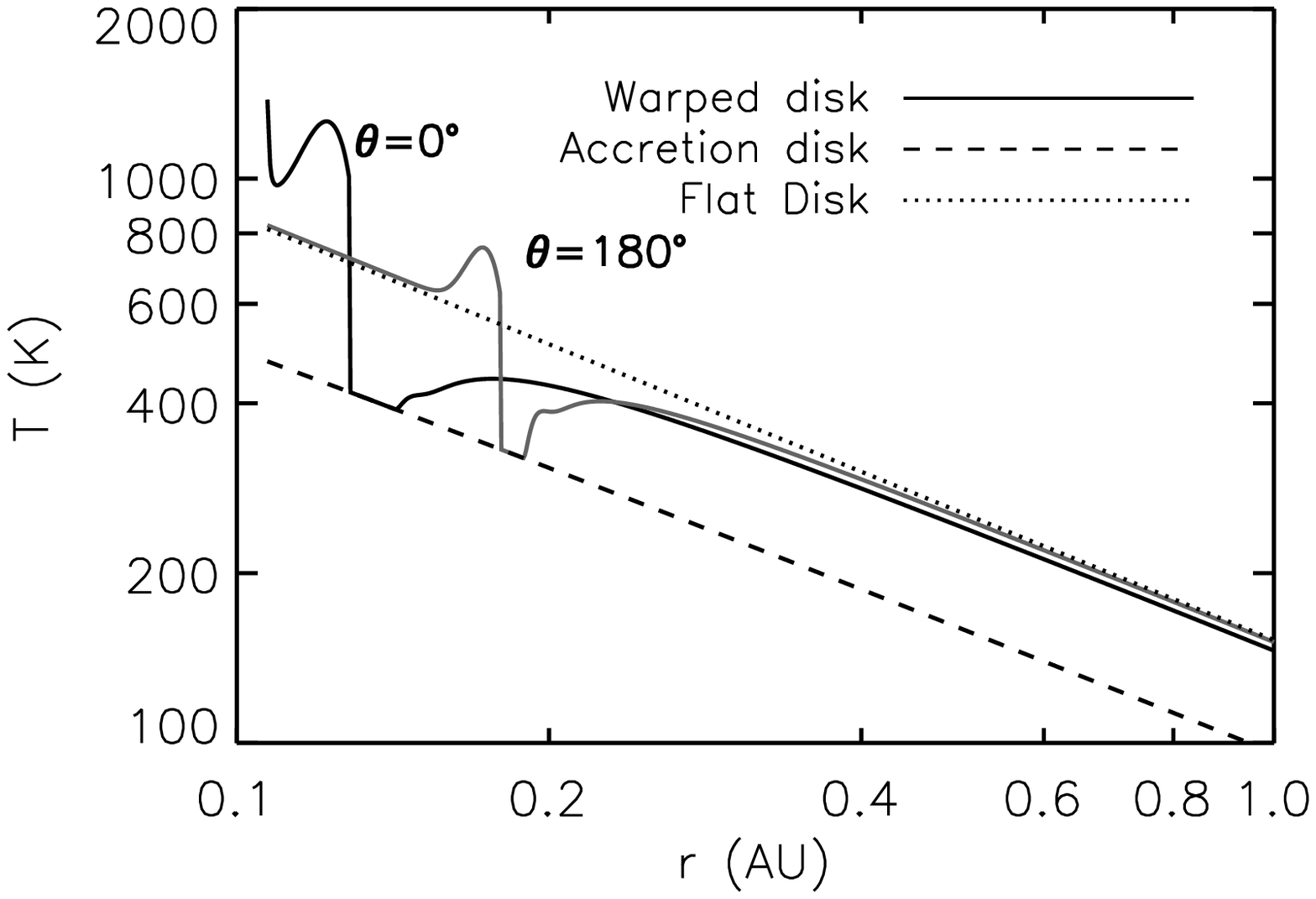}
\includegraphics[width=0.33\textwidth]{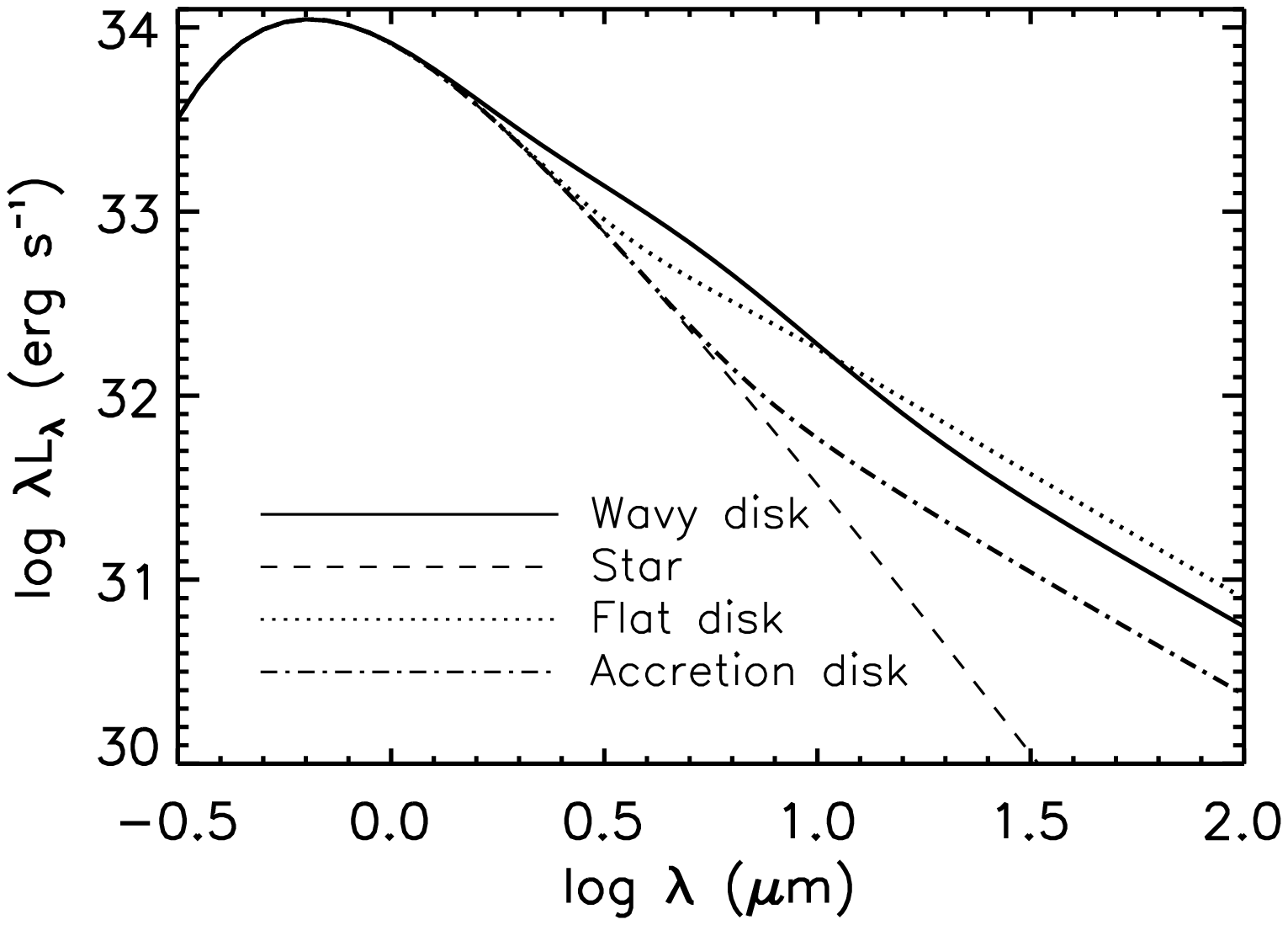}
\caption{Spiral wave disk. The left and middle panels show the temperature distribution for a spiral wave with $g=0.05$ (left panel) and with $g=0.01$ (middle panel). The dark and light solid lines are cuts at $\theta=0,180^{\circ}$ respectively. The dotted and dashed lines are temperature structures for a flat passive and flat accretion disk respectively. On the right is the SED of a typical disk with a spiral wave ($g=0.05,i=60^{\circ},\alpha=0^{\circ}$).  \label{warp3}}
\end{figure}

\begin{figure}
\plottwo{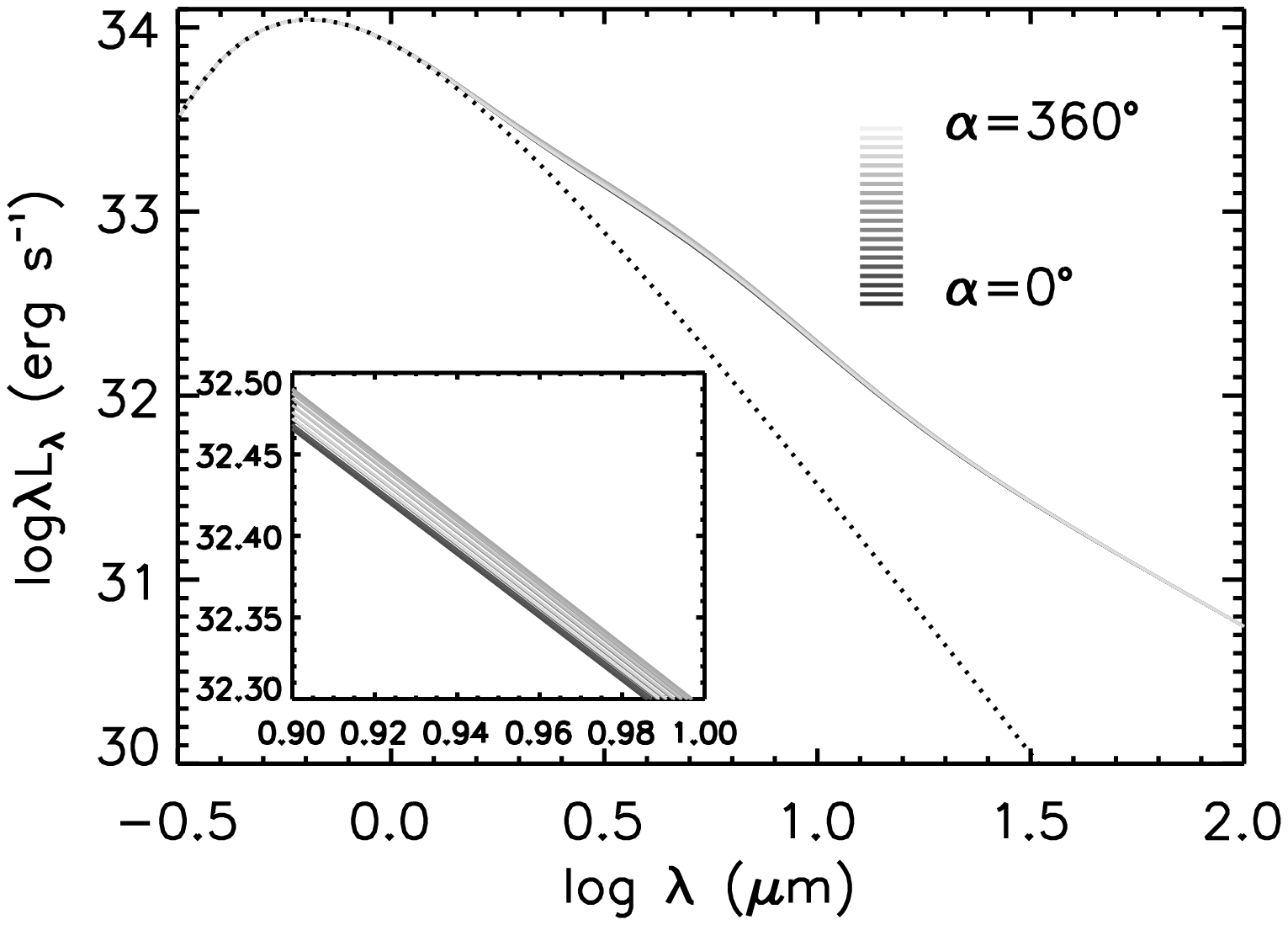}{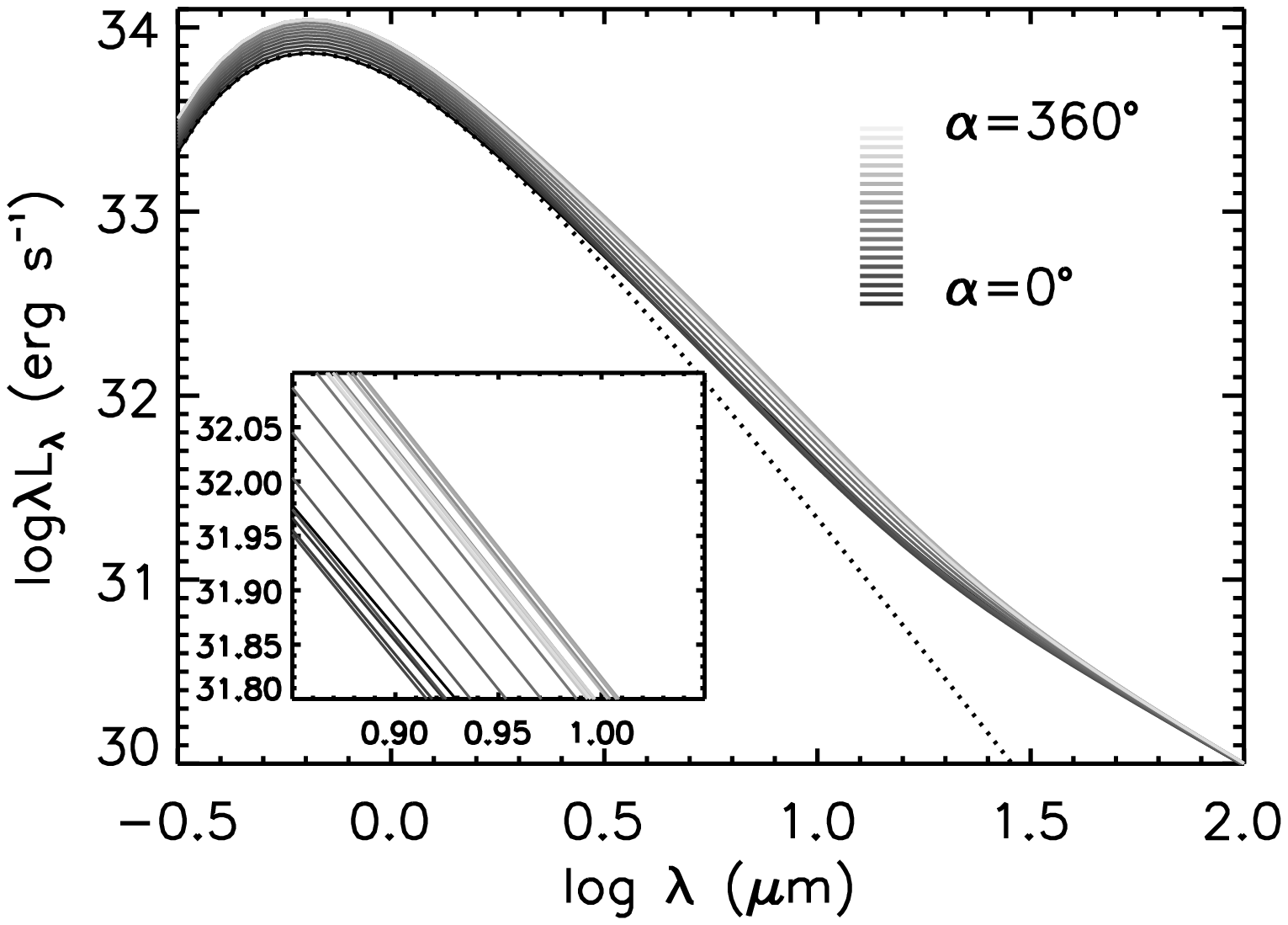}
\caption{SEDs of a precessing spiral wave with $g=0.05$ seen at $i=60,85^{\circ}$ (left and right panels respectively). From dark to light solid lines the disk rotates from $\alpha=0^{\circ}$ to $\alpha=360^{\circ}$. Viewing the spiral wave from different orientations has a small change on the infrared flux, unless the wave occults the star. \label{warp3_prec}}
\end{figure}

\begin{figure}
\plotone{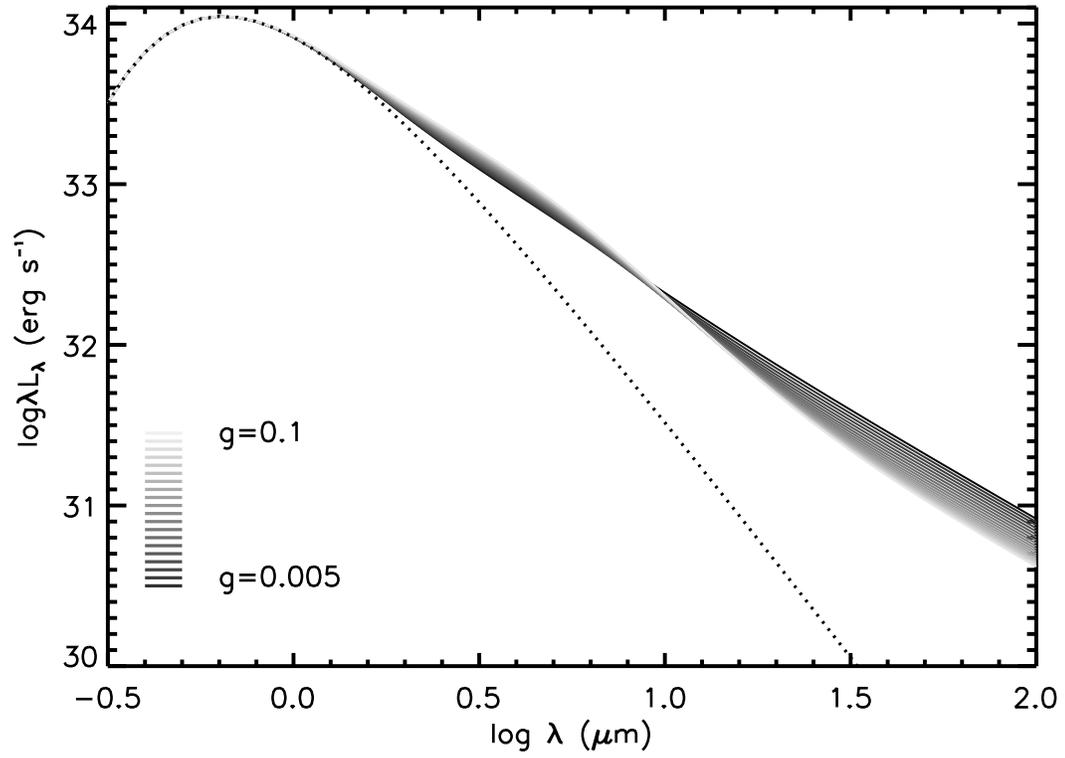}
\caption{SEDS for a growing spiral wave. The disk is viewed at $i=60^{\circ},\alpha=180^{\circ}$ and the models vary $g$ from 0.005 to 0.1. The change in the far-infrared flux is much larger than the change in the mid-infrared flux. \label{warp3_grow_i60}}
\end{figure}

\begin{figure}
\plotone{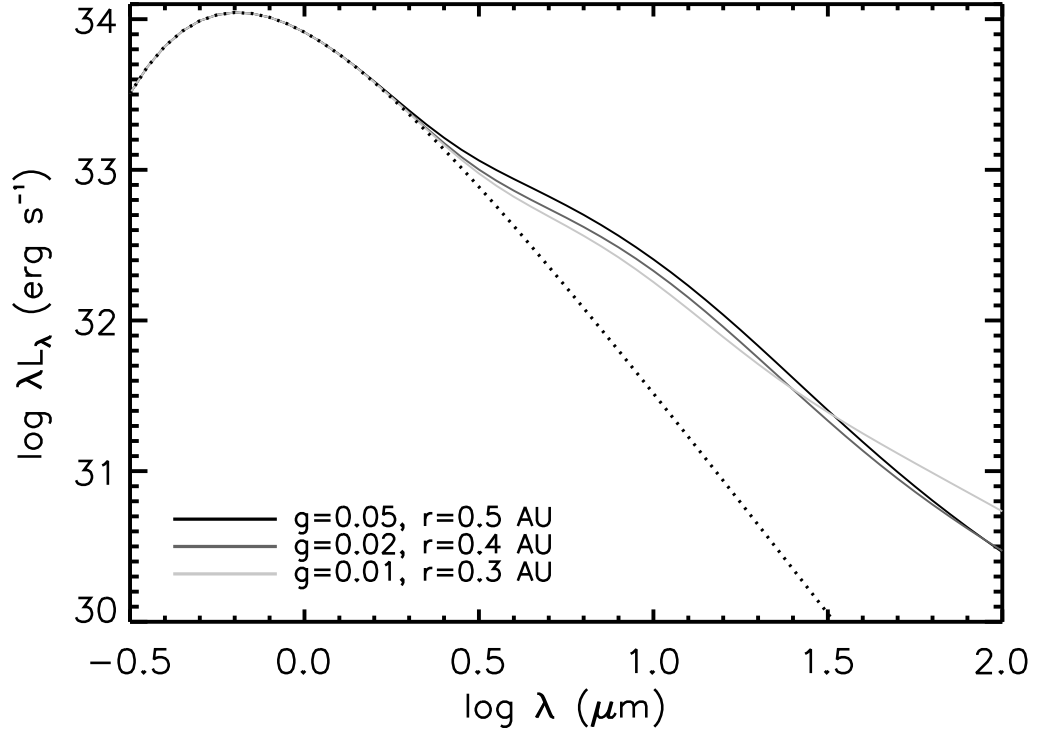}
\caption{SEDs for the thermal instability wave. The three models represent g=0.05 at r=0.5 AU, g=0.02 at r=0.4 AU and g=0.01 at r=0.1 AU from dark to light lines. The photosphere is shown for comparison.\label{l31_warp3_ti}}
\end{figure}

\clearpage

\begin{figure}
\plotone{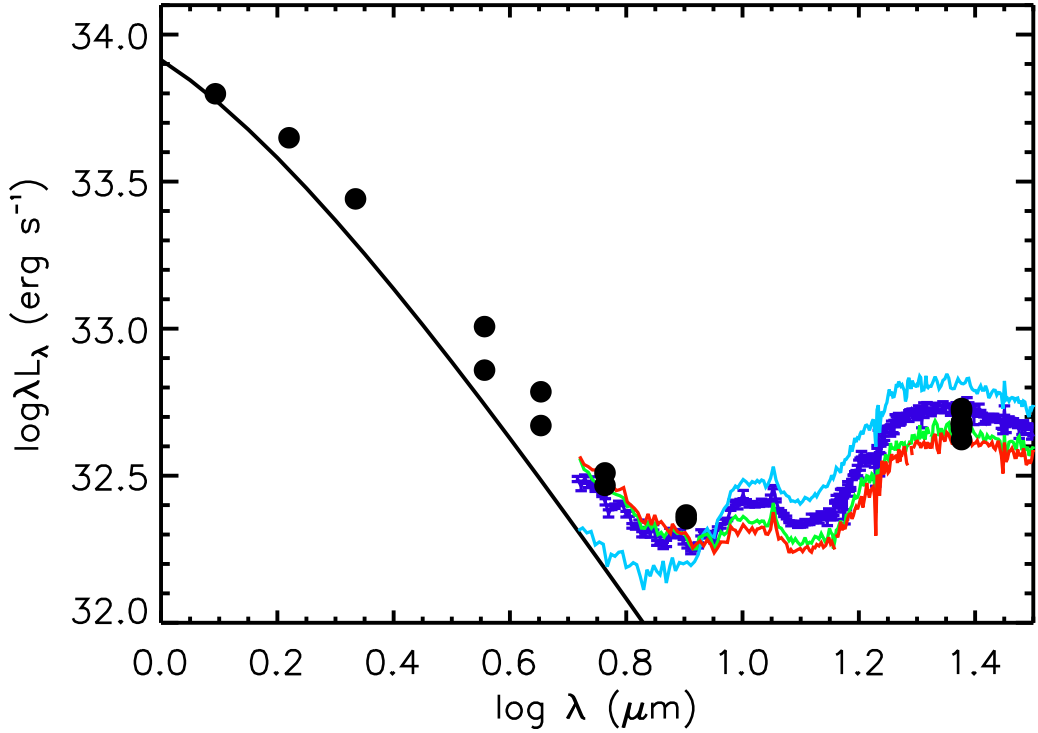}
\caption{Observations of LRLL 31. Black circles are the photometry from 2MASS, two epochs of IRAC photometry and 10 epochs of MIPS 24\micron\ photometry. Colored lines are the four IRS spectra of LRLL 31. The dark blue and light blue were taken in October 2007 and were separated by one week. The green and red spectra were taken in February/March 2008 and were separated by one week. Substantial variability is seen throughout the mid-infrared. Error bars on the dark blue spectra show the uncertainty in the spectra, which are similar to the uncertainties in the other spectra, and are smaller than the variability in the flux. The dark solid line is a 5700 K blackbody representing the stellar photosphere for comparison, and the observations have been scaled to match the photospheric luminosity at J band, assuming the J band flux is completely photospheric.\label{l31obs}}
\end{figure}

\begin{figure}
\plottwo{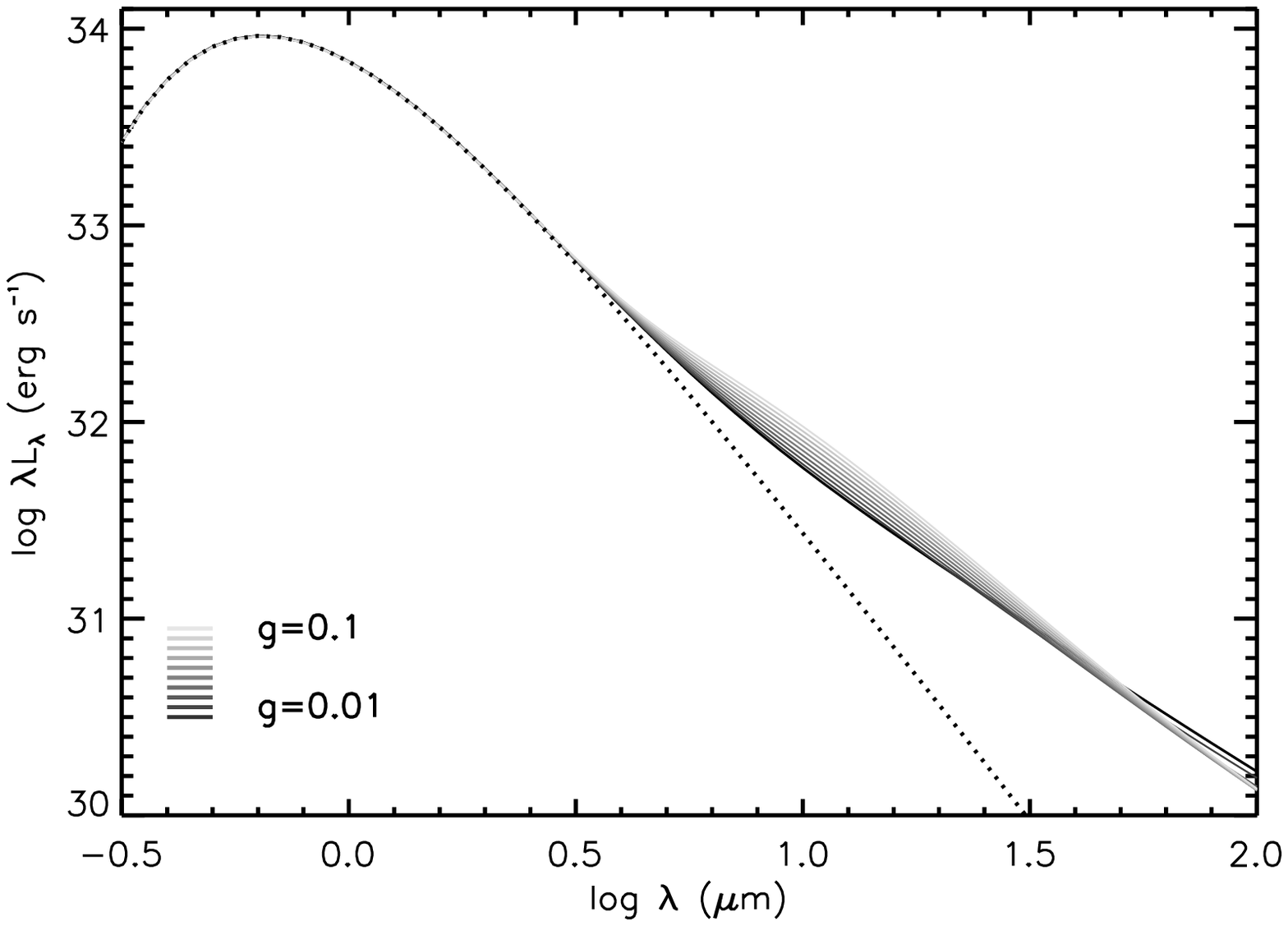}{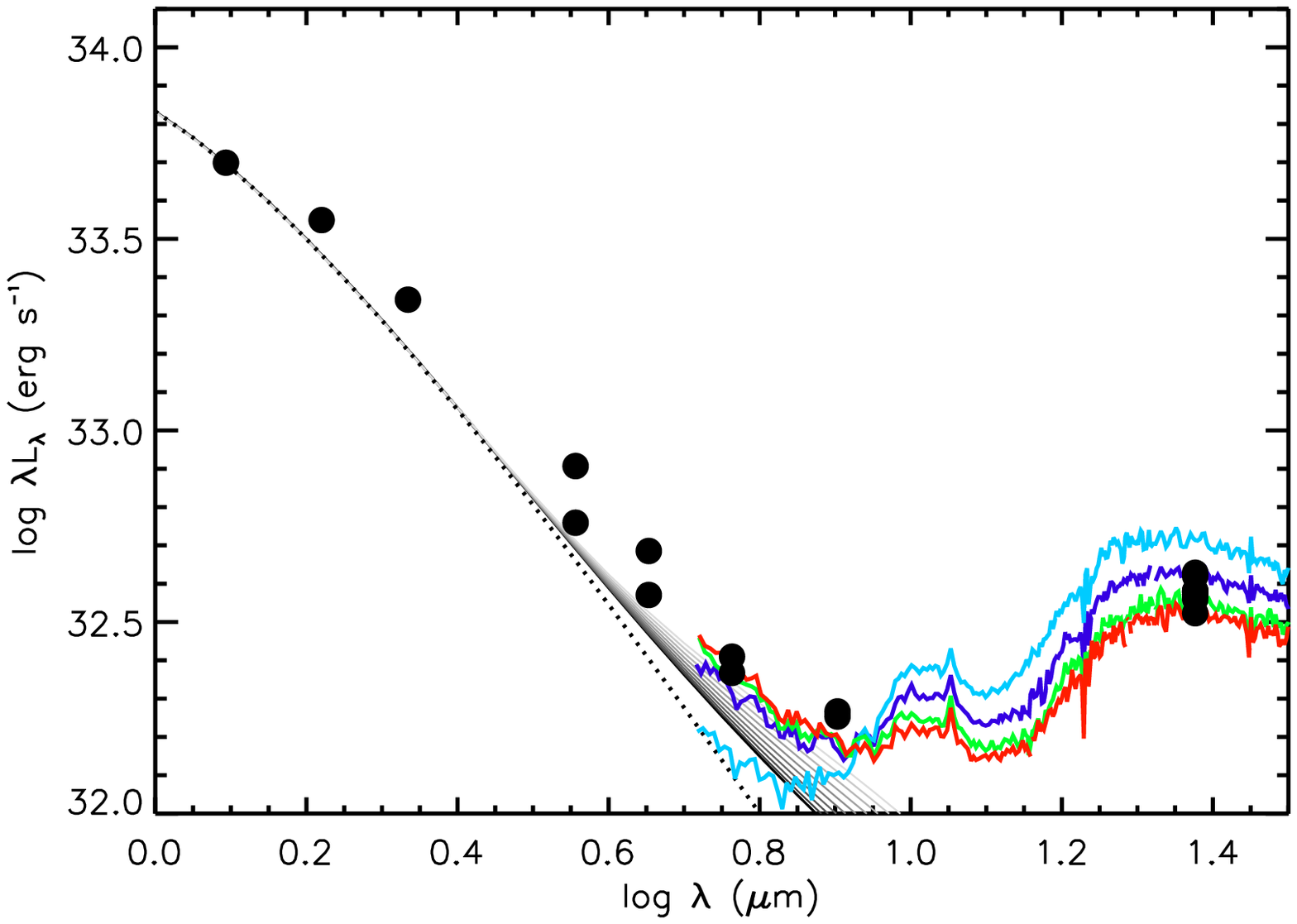}
\caption{SEDs for a growing middle warp ($g=0.01-0.1$ from dark to light lines) viewed at $i=85^{\circ},\alpha=180^{\circ}$. On the left is the entire model, while on the right is a close-up of the model with the observations. The observations have been scaled to match the J band flux of the model, assuming the J band observation is completely photospheric.\label{l31_warp_grow}}
\end{figure}

\begin{figure}
\includegraphics[width=0.5\textwidth]{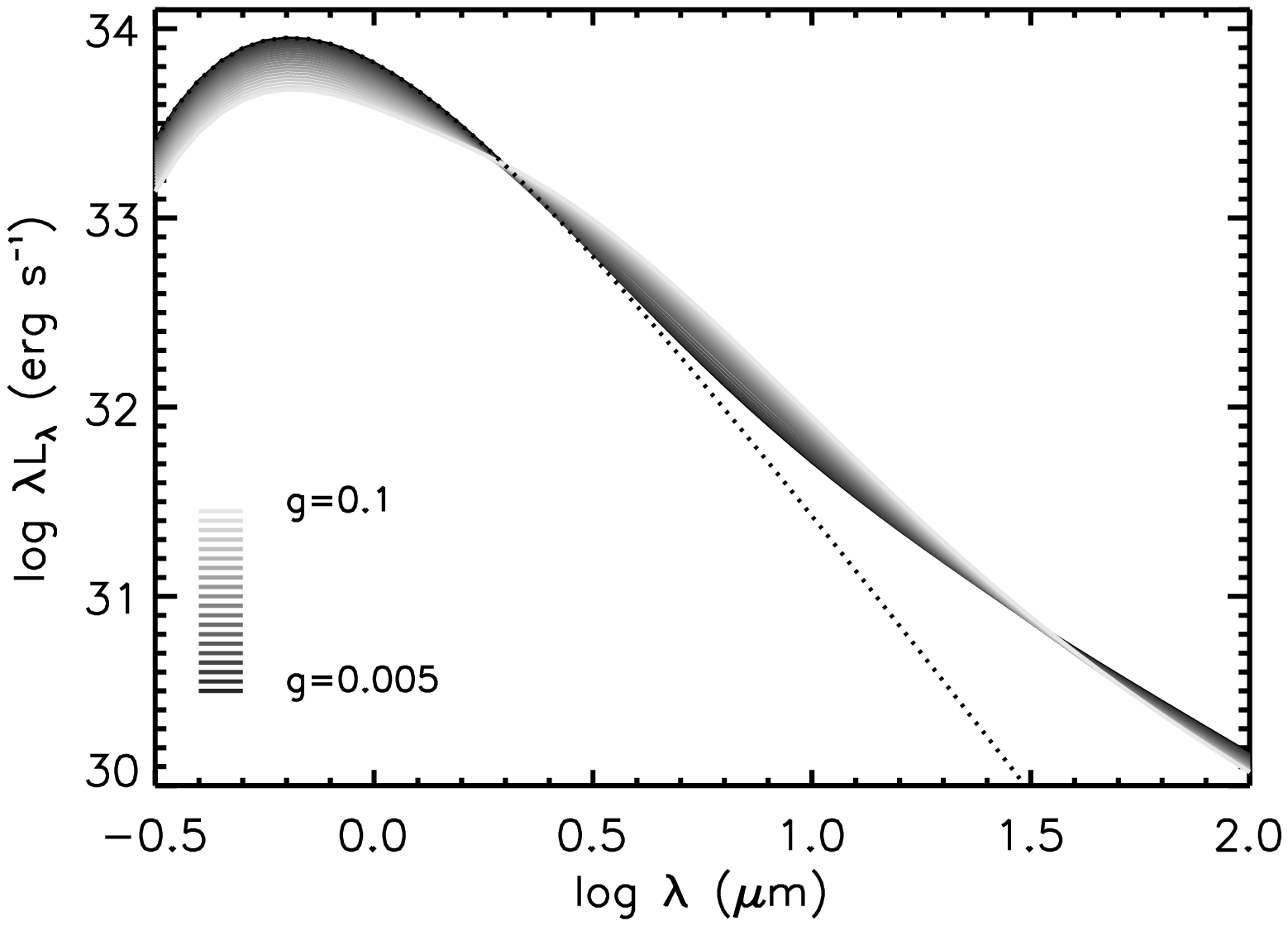}
\includegraphics[width=0.5\textwidth]{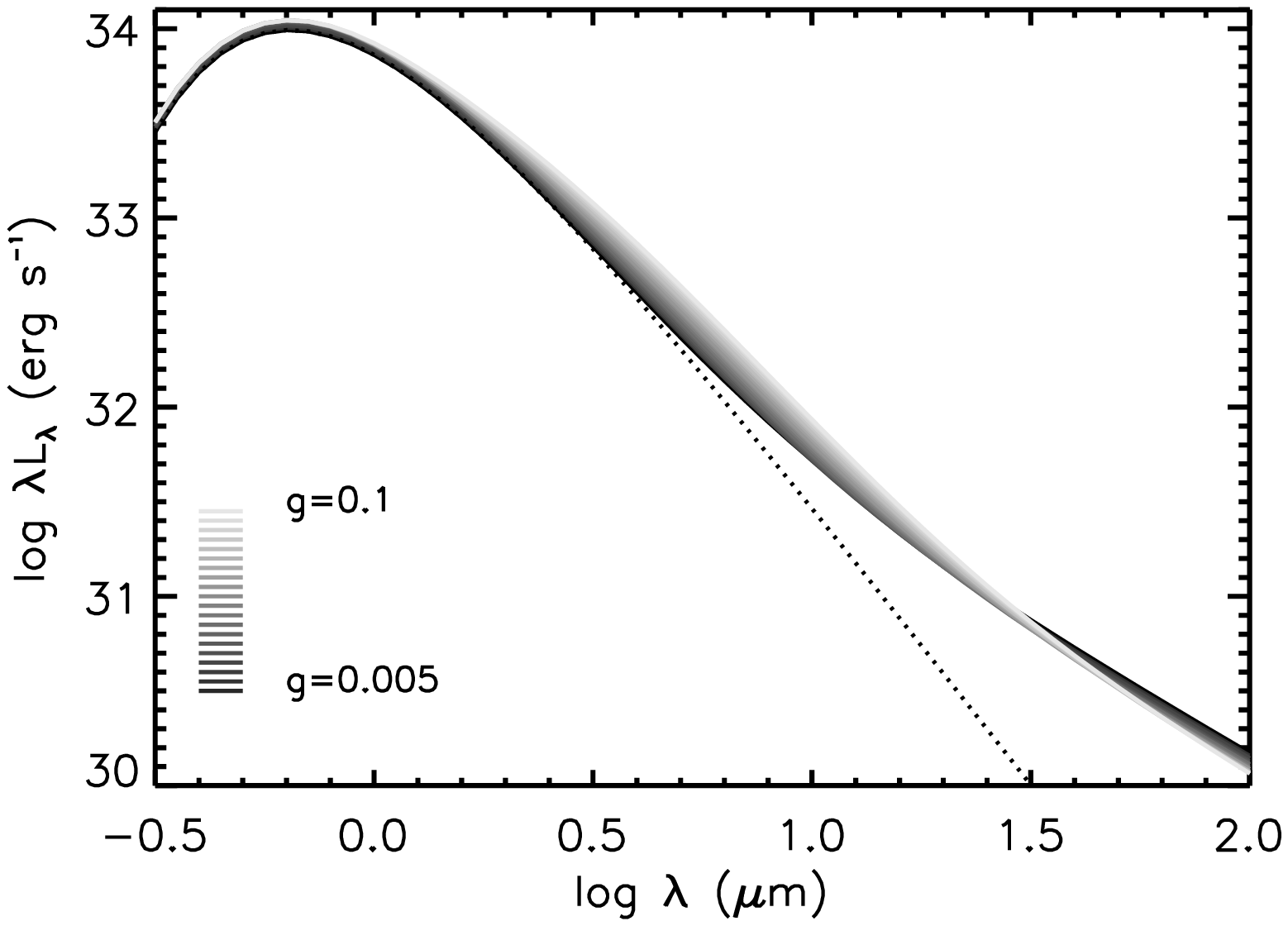}
\hfill
\includegraphics[width=0.5\textwidth]{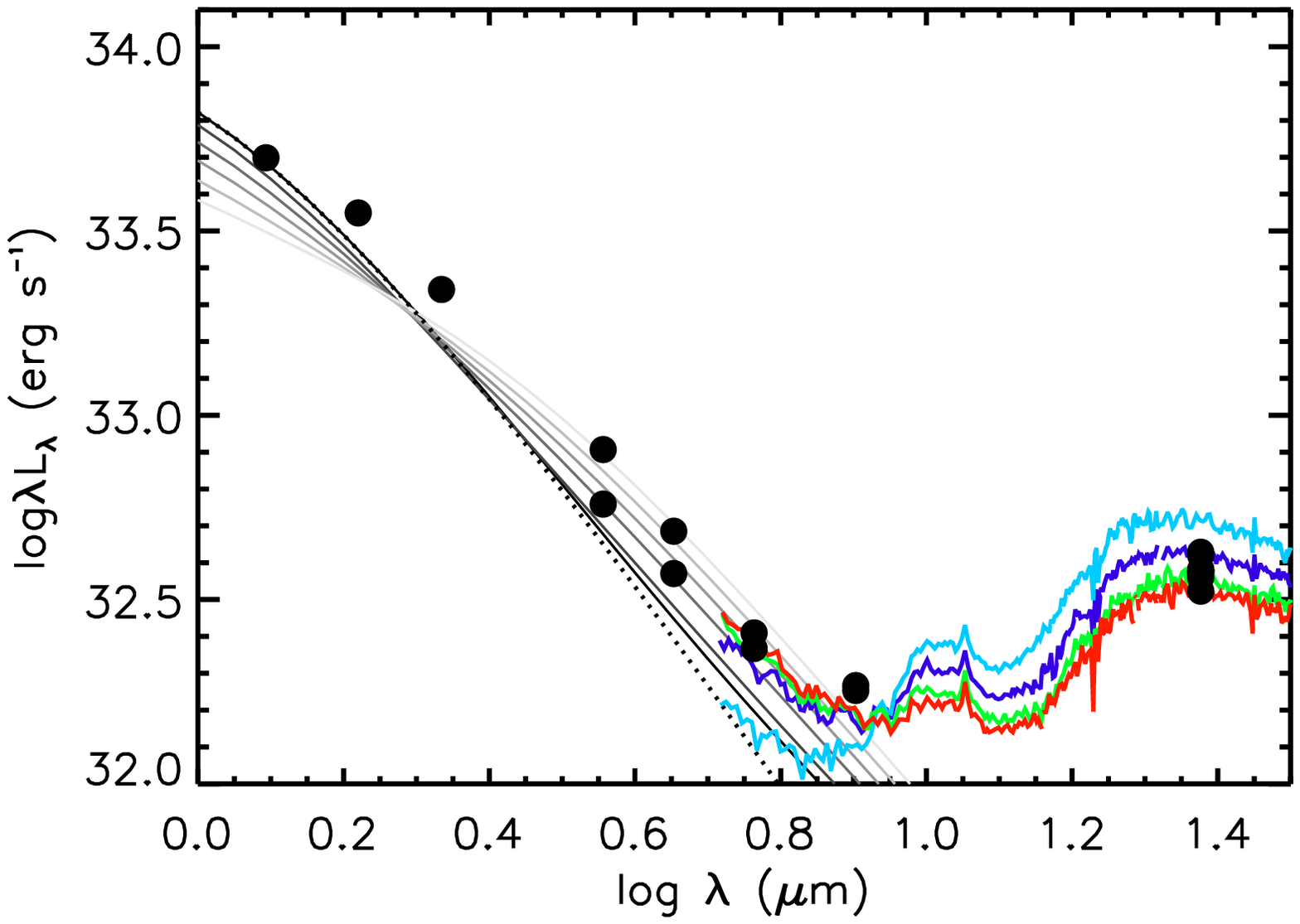}
\includegraphics[width=0.5\textwidth]{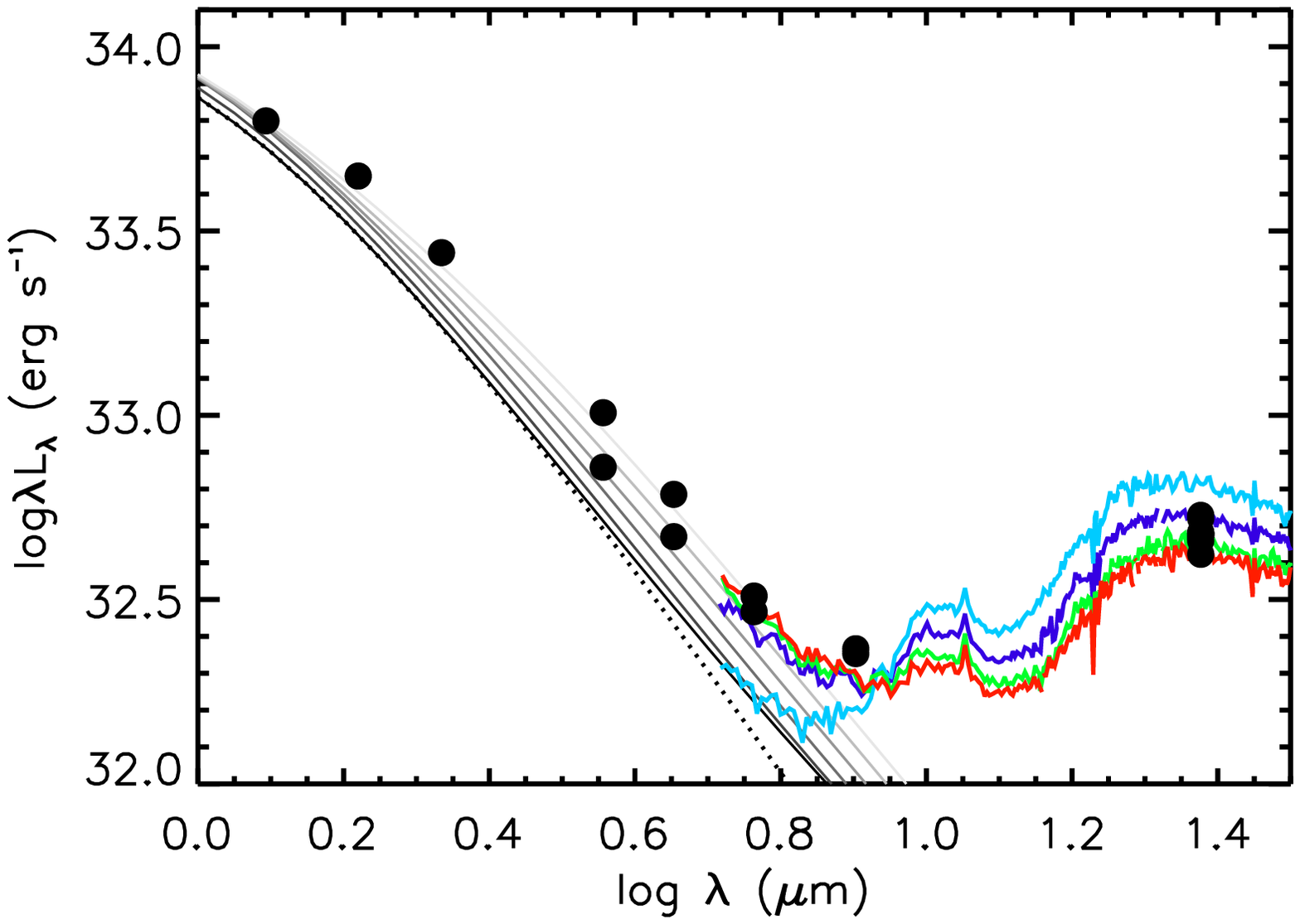}
\caption{SEDs of LRLL 31 with a growing inner warp. The models vary $g$ from 0.005 to 0.1 (from dark to light lines). The models on the left are seen at an inclination of $i=85^{\circ}$, while on the right are the models seen at $i=95^{\circ}$. The top panels show the entire models while the bottom panels zoom in and include the observations. Only models for $g=0.005,0.02,0.04,0.06,0.08,0.1$ are shown for clarity. The observations have been scaled to match the J band flux of the model, assuming the J band observation is completely photospheric. \label{l31_warp2_grow}}
\end{figure}

\begin{figure}
\plottwo{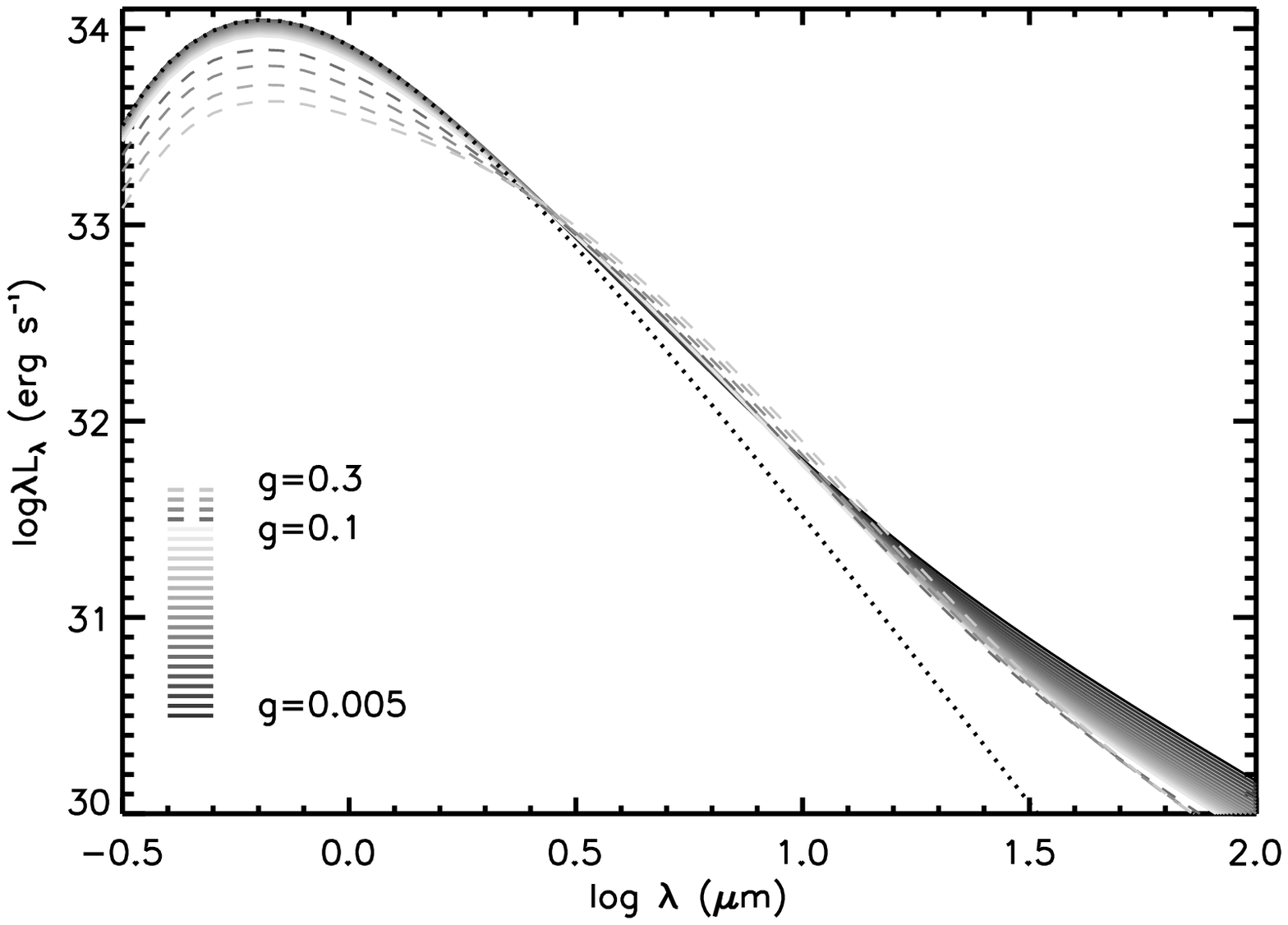}{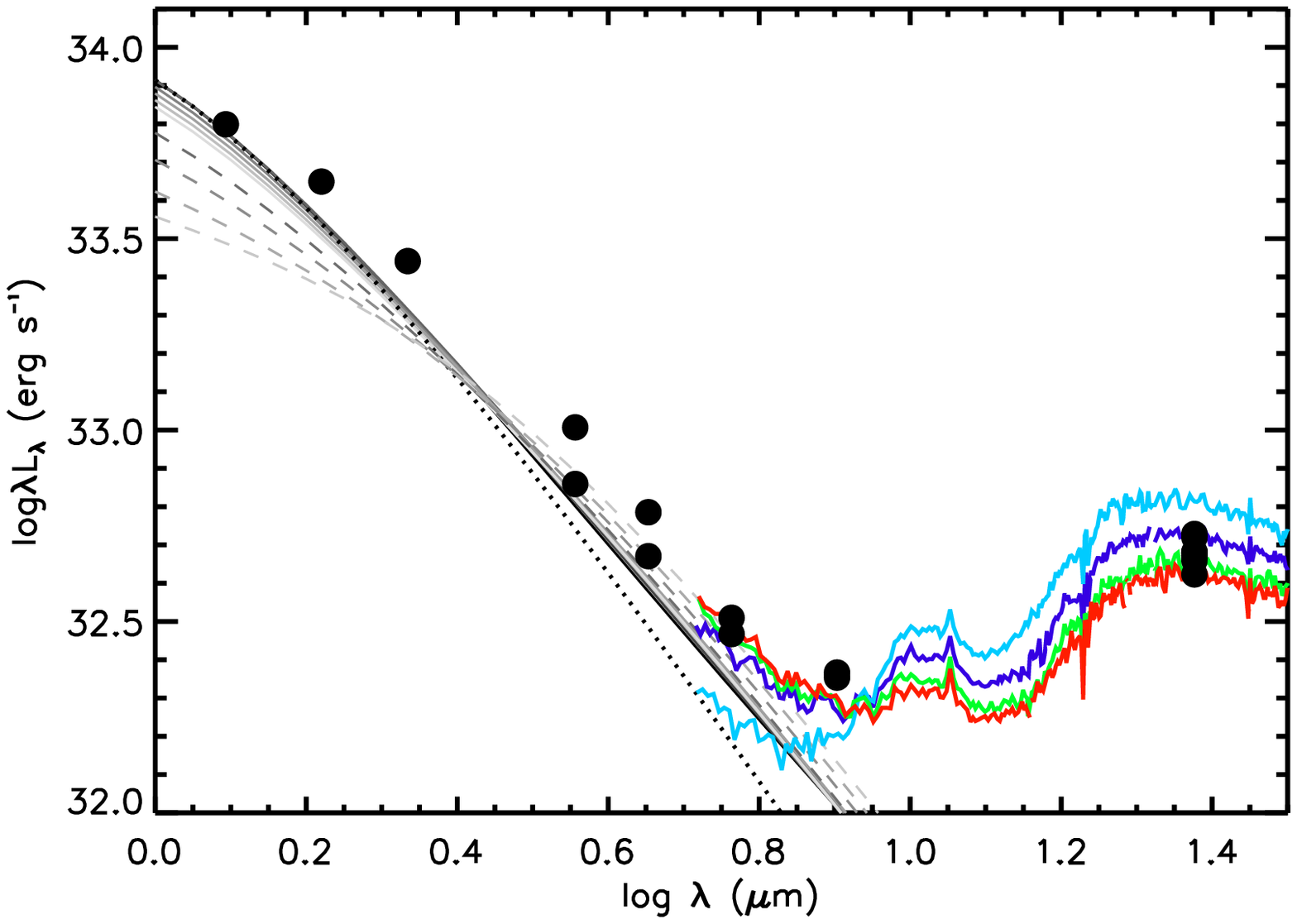}
\caption{SEDs for a growing spiral wave around LRLL 31. The disk is viewed at $i=85^{\circ},\alpha=180^{\circ}$. The models vary $g$ from 0.005 to 0.1. The dashed lines are models with $g=0.15,0.2,0.25,0.3$, larger than previously used for the inner warp. On the right is a close up of the observations along with the models for $g=0.005,0.02,0.04,0.06,0.08,0.1,0.15,0.2,0.25,0.3$. The observations have been scaled so that the J band flux matches the models, assuming the J band observation is completely photospheric. \label{l31_warp3_grow}}
\end{figure}

\begin{figure}
\plotone{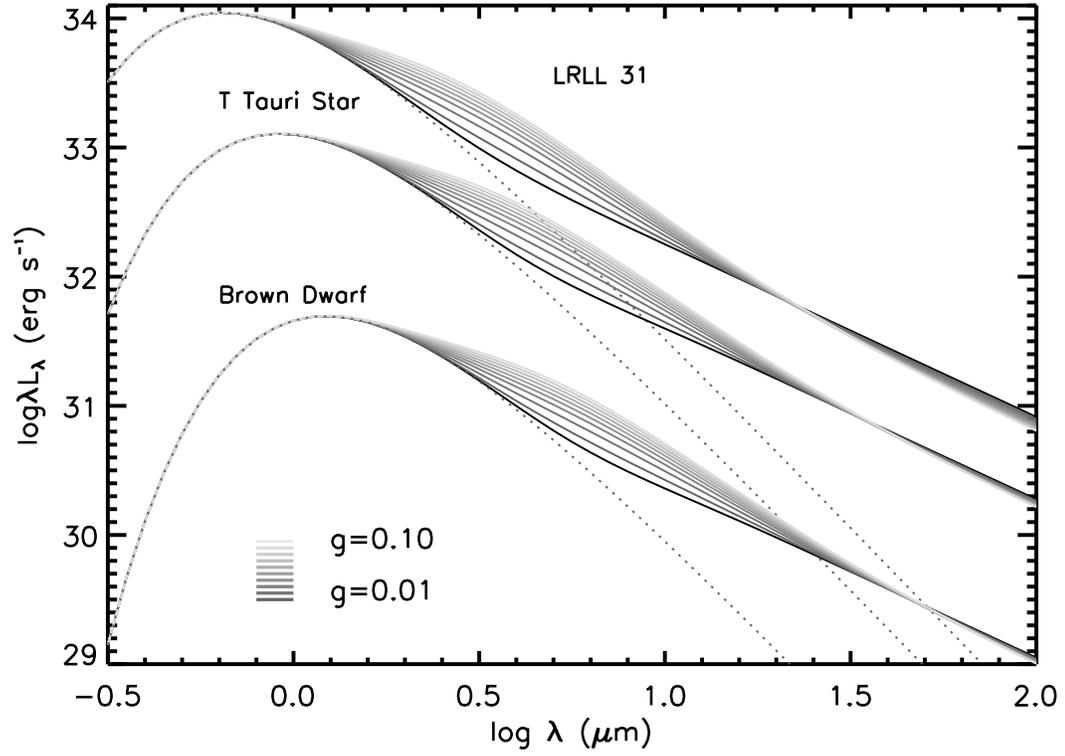}
\caption{SEDs for an inner warp surrounding LRLL 31, a typical T Tauri star and a brown dwarf seen at $i=60^{\circ}$. For each star the models vary $g$ from 0.01 to 0.1(dark to light grey solid lines) and the stellar photosphere is shown with the dotted line. The change in mid-infrared flux does not depend strongly on the central star. \label{tts}}
\end{figure}

\clearpage




\begin{center}
{\bf APPENDIX}
\end{center}

\section{Outer Warp + Flat Extension}
\subsection{Adding $\alpha$ Dependence to Outer Warp}
TB96 describe how to calculate the SED for an outer warp seen at various inclination angles. Their general method for calculating the SED also includes a dependence on azimuthal viewing angle, although their detailed treatment of various occultation effects (the star blocking the far side of the disk, the disk blocking the star, etc.) does not include this dependence. Since these disks are non-axissymetric the SED can depend substantially on the azimuthal viewing angle, $\alpha$, of the observer. In this section we describe how we have added this dependence into the equations of TB96. We do not include a detailed description of the derivation of the equations, but merely state most of them and some of the geometric logic behind their modification.

The first modification is to equation (6) of TB96, which describes the calculation of the flux from the disk based on the temperature of the disk. This equation assumes that the disk is viewed along $\alpha=0$ so that the upper disk from 0 to $\pi/2$ looks the same as the upper disk from $3\pi/2$ to $2\pi$. When $\alpha\neq0$ this symmetry is broken and the individual components of the disk must be considered. While the temperature of a concave, or convex, piece of the disk does not change with viewing angle, the orientation of each of the concave, or convex pieces, changes and must be treated separately. For the outer warp this becomes:

\begin{eqnarray}
\textstyle
F_{\nu,{\bf u}} = \int^{R_{disk}}_{r_{min}}\left[\int_0^{\pi/2}B(T_{concave})f_{up}+\int_0^{\pi/2}B(T_{convex})f_{low}+\int_{\pi/2}^{\pi}B(T_{concave})f_{low}+\int_{\pi/2}^{\pi}B(T_{convex})f_{up}\right]\nonumber\\
\textstyle
+\int^{R_{disk}}_{r_{min}}\left[\int_{\pi}^{3\pi/2}B(T_{convex})f_{up}+\int_{\pi}^{3\pi/2}B(T_{concave})f_{low}+\int_{3\pi/2}^{2\pi}B(T_{concave})f_{up}+\int_{3\pi/2}^{2\pi}B(T_{convex})f_{low}\right]
\end{eqnarray}

The next change comes in the appendix to the functional form of the parameter $C$. The function $C$ is used to define the points that are along the line of sight with the star. If the line intersects the star then we need to worry about whether the disk blocks the star or the star blocks the disk. If this line does not intersect the star then the disk cannot block the star and the star cannot block the disk. The definition of $C$ changes from $C=r\sin\theta$ to $C=r\sin (\theta-\alpha )$. Also the radial part of the deformation used in equation (A6) of TB96 is taken to be $H(r)=gR_{disk}\left(\frac{r}{R_{disk}}\right)^n\cos\alpha$. For $\alpha=\pi/2$ the disk along the line of sight is flat and the radial part of the height will remain at zero, while along $\alpha=\pi$ the disk curves below the midplane as expected.

The final change comes when calculating the stellar flux. In equation (A12) of TB96 we take $h(r,\theta)$ to be $h(r,\alpha)$ since this represents the part of the disk that will block the star. As the azimuthal angle increases the disk blocks less of the star because the height of the disk is smaller. We make more changes to how the stellar flux is calculated, which are described below, but when it comes to the occultation of the star by the disk this is the only change.

In the end we are able to run our models from $0<i<\pi$ and $-\pi/2<\alpha<\pi/2$. Due to the symmetry of the disk this covers all possible viewing angles allowing us to accurately model the precession of the warp, as well as observe the warp from an arbitrary angle. 

\subsection{Flat extension of Outer Warp}
We have taken the outer warp model and added a flat extension beyond it in order to treat disks where the warp is not at the outer edge of the disk. The warp will shadow the outer disk, changing its temperature structure. For simplicity we assume that the outer disk is a flat blackbody. The temperature can be derived using the same formula as with the warped disk (Equation 6), but with different definitions for the integration boundaries. Half of the flat extension will be shadowed while half will not be shadowed. For the side of the flat extension that is beyond the part of the warp the goes below the midplane there is no shadowing of the disk and the integration ranges over:
\begin{eqnarray}
\varepsilon_{min}=0\nonumber\\
\varepsilon_{max}=\pi/2\nonumber\\
\\
\delta_{min}=0\nonumber\\
\delta_{max}=\arcsin(R_*/d)\nonumber\\
d^2=r^2+h^2\nonumber\\
\end{eqnarray}

For the part of the flat extension that lies behind the warp that stretches above the midplane the definition of $\delta_{min}$ changes. In this case $\delta_{min}$ is set by the angle between the warp and the point $P(r,\theta)$ in the disk.

\begin{equation}
\delta_{min}=\arctan\left(\frac{gR_{warp}\cos(\theta)}{(r-R_{warp})}\right)
\end{equation}

This takes into account shadowing of the flat extension to the disk due to the warp. Once the temperature structure has been determined the flux can be derived using the equation for the flux from a disk (Equation 5). 

\section{Inner Warp}
\subsection{Temperature Profile of the Inner Warp}
In this section we describe the method for calculating the SED of a disk with an inner warp. In the text we laid out the basic equations from TB96 that are needed to calculate the temperature structure. As mentioned in the text the essential difference between the inner warp and outer warp comes in calculating $\delta_{max},\delta_{min},\varepsilon_{min},\varepsilon_{max}$ for each point $P(r,\theta)$, which are used in equation 6. From here the disk is split into two sides that are treated separately. The convex side is the side of the disk that faces the star on the inner edge and receives the most direct heating from the star. For this side, the integration ranges over:

\begin{eqnarray}
\varepsilon_{min}=0\nonumber\\
\varepsilon_{max}=\pi/2\nonumber\\
\\
\delta_{min}=-\arctan(\partial h/\partial r)\nonumber\\
\delta_{max}=\arcsin(R_*/d)\nonumber\\
d^2=r^2+h^2\nonumber\\
\end{eqnarray}

\begin{figure}
\epsscale{.75}
\plotone{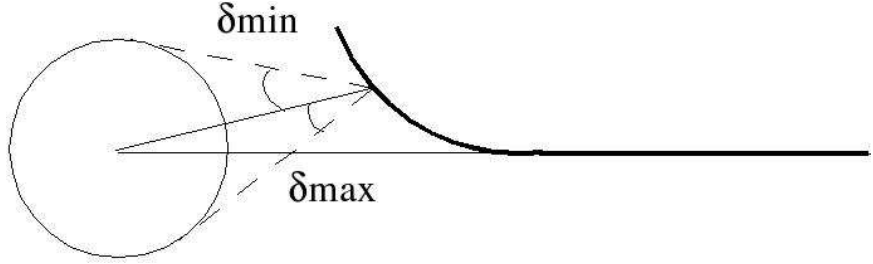}
\caption{Schematic diagram of the limits on $\delta$ for the convex side of the disk.\label{del_vex}}
\end{figure}

\begin{figure}
\epsscale{.75}
\plotone{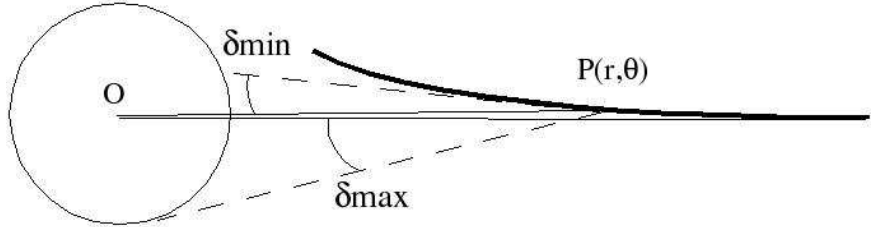}
\caption{Schematic diagram of the limits on $\delta$ for the convex side of the disk. In this case the star is blocked by some of the disk. The shallow slope of the disk prevents all of the star from being seen. \label{delblock_vex}}
\end{figure}

Figure~\ref{del_vex} demonstrates the limits on $\delta$ for the convex side of the disk. The definition of $\delta_{min}$, demonstrated in figure~\ref{delblock_vex}, comes from the inner disk blocking light from the top of the star. The inner disk will limit the field of view of the point $P(r,\theta)$ as it looks toward the star. Traveling out from the star, less of the star will be seen by the disk because the shallower slope of the disk will cause more of the star to be blocked. In the limit of a flat disk far from the star $\delta_{min}$ approaches zero and the disk can only see half of the star. If the point $P(r,\theta)$ on the disk is close enough to the star then the disk can see all of the star and $\delta_{min}=-\delta_{max}$. The limits on $\varepsilon$ assume that the scale height of the disk does not change across the face of the disk, which will be an accurate approximation far from the star. 

The concave side of the disk is the side that does not directly face the star. Since it does not face the star much of the inner parts of the disk will be blocked by the warp and will only be heated by viscous dissipation. The condition for the point $P(r,\theta)$ on the concave part of the disk to see any of the star is:

\begin{equation}
\frac{h(r_{min},\theta)}{r-r_{min}}<\frac{R_*}{r}
\end{equation}

\begin{figure}
\epsscale{.75}
\plotone{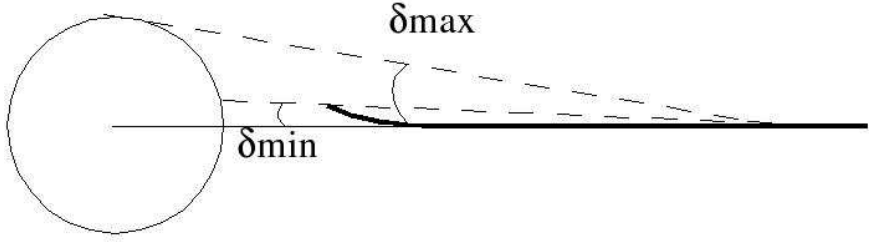}
\caption{Schematic diagram of the limits on $\delta$ for the concave side of the disk.\label{delmin_cave}}
\end{figure}

If the point $P(r,\theta)$ meets this condition then this point can see some of the star and $\delta_{min}$ becomes (fig~\ref{delmin_cave})

\begin{equation}
\delta_{min}=\arctan(\frac{h(r_{min},\theta)}{r-r_{min}})
\end{equation}

The rest of the limits stay the same. In the limit of a perfectly flat disk $\delta_{min}$ approaches 0 and the point $P(r,\theta)$ is irradiated by only half of the star. For a large warp the only heating by this side of the disk will be from viscous dissipation because the warp will block the star over most of the disk. 

\subsection{Calculating the SED for an Inner Warp}
In this section we describe the procedure for converting the temperature structure into a spectral energy distribution (SED). From TB96, the flux emitted by the disk is

\begin{equation}
F_{\nu,{\bf u}}= \int\int_{disk surface} B_{\nu}(T(r,\theta))dS{\bf n_d}\cdot{\bf u}
\end{equation}

In this case {\bf u} is the vector along the line of sight to the observer from the center of the star and ${\bf n_d}$ is the normal to the disk at the point $P(r,\theta)$. The vector ${\bf u}$ can be defined in terms of the azimuthal and polar angles to the line of sight, $\alpha$ and $i$ respectively. The area of the disk along the line of sight is given by

\begin{equation}
\textstyle
dS {\bf n_d}\cdot{\bf u} = r\left[\left(\frac1{r}\frac{\partial h}{\partial \theta}\sin\theta-\frac{\partial h}{\partial r}\cos\theta\right)\cos\alpha\sin i-\left(\frac1{r}\frac{\partial h}{\partial\theta}\cos\theta+\frac{\partial h}{\partial r}\sin\theta\right)\sin\alpha\sin i+\cos i\right]drd\theta
\end{equation}

The angle $\alpha$ ranges from $-\pi$/2 to $\pi$/2 while the inclination $i$ ranges from 0 to $\pi$. This covers all possible viewing angles of the disk, since the symmetry of the disk makes some viewing angles redundant. 

Splitting up the equation for the flux from the disk helps to make the problem simpler to understand and more tractable. It also fits with the fact that we do not need to calculate the temperature structure of the entire disk. The symmetry of the disk allows us the calculate the temperature of the convex and concave side from $0<\theta<\pi/2$ and then apply this temperature profile to the rest of the disk. The integral is split into eight parts:

\begin{eqnarray}
\textstyle
F_{\nu,{\bf u}} = \int^{R_{disk}}_{r_{min}}\left[\int_0^{\pi/2}B(T_{concave})f_{up}+\int_0^{\pi/2}B(T_{convex})f_{low}+\int_{\pi/2}^{\pi}B(T_{concave})f_{low}+\int_{\pi/2}^{\pi}B(T_{convex})f_{up}\right]\nonumber\\
\textstyle
+\int^{R_{disk}}_{r_{min}}\left[\int_{\pi}^{3\pi/2}B(T_{convex})f_{up}+\int_{\pi}^{3\pi/2}B(T_{concave})f_{low}+\int_{3\pi/2}^{2\pi}B(T_{concave})f_{up}+\int_{3\pi/2}^{2\pi}B(T_{convex})f_{low}\right]\label{eqn_flux}
\end{eqnarray}

where $f(r,\theta)=dS {\bf n_d}\cdot{\bf u}p(r,\theta)$, and $p(r,\theta)$ is a binary function used to determine if the point $P(r,\theta)$ is visible to the observer. The integration is done over both the upper and lower sides of the disk in order to account for inclination angles greater than $90^{\circ}$ where the lower half of the disk is visible. If the inclination is 0 then the observer is face on to the upper half of the disk, which has both a concave and convex side. If the inclination is $180^{\circ}$ then the observer is face on to the lower half of the disk, which includes both a concave and a convex side. Treating each quarter of the disk separately allows us to use the symmetry of the temperature profile but still treat general azimuthal viewing angles.

\subsubsection{Calculating the value of p}

The above description sets out the basics for how to calculate the temperature structure and SED for a warped inner disk. Most of this is derived from TB96, which treated these situations generally enough to apply to any type of warp. The main differences between this inner warp and the outer warp from TB96 comes from the calculation of $p(r,\theta)$. This section describes the conditions used to calculate $p(r,\theta)$ for the particular warp used here.

The first condition is that the observer is facing the point $P(r,\theta)$. For inclinations less than $90^{\circ}$ the observer will see mostly the upper half of the disk, while at inclinations greater than $90^{\circ}$ the observer will see mostly the lower half of the disk. There are select inclinations close to edge on where at inclinations less than $90^{\circ}$ some of the lower disk can be seen. For example, if figure~\ref{delblock_vex} had an observer in the upper left viewing the disk close to edge on they would be able to see some of the lower convex side that is illustrated in the figure. In general it can be determined if the observer is facing the point $P(r,\theta)$ based on the dot product ${\bf n_d}\cdot{\bf u}$. The normal is defined as extending on the upper side of the disk and the dot product will be greater than zero if ${\bf n_d}$ and ${\bf u}$ lie along the same direction. Therefore, the upper part of the disk can be seen if the dot product is greater than 0 while the lower part of the disk can be seen when the dot product is negative.

\begin{figure}
\epsscale{.75}
\plotone{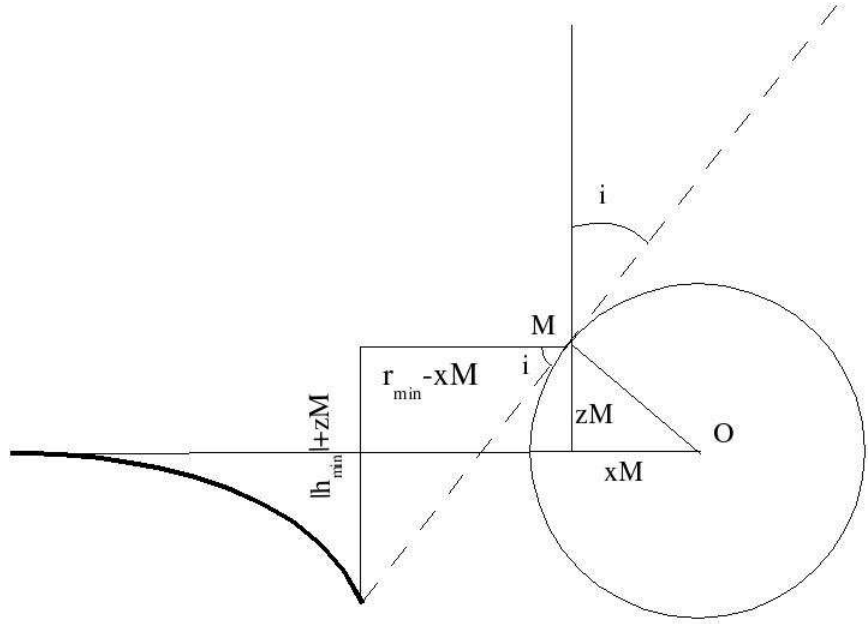}
\caption{Schematic diagram demonstrating when the star starts to block some of the disk. \label{inclim}}
\end{figure}

Now we determine if the point $P(r,\theta)$ is blocked by either the star or the disk. First we consider whether the star blocks the far side of the disk. This applies for inclinations less than $90^{\circ}$ where the star may block part of the upper convex side, as is demonstrated in figure~\ref{inclim}. The limit at which this condition applies is given by

\begin{equation}
\tan i_{lim}=\frac{r_{min}-R_*\cos i_{lim}}{gr_{min}\cos\alpha+R_*\sin i_{lim}}
\end{equation}

\begin{deluxetable}{cccc}
\tablewidth{0pt}
\tablecaption{Limiting inclination for occultation of disk by star\label{ilim}}
\tablehead{\colhead{g}&\colhead{$r_{min}=2R_*$}&\colhead{$r_{min}=5R_*$}&\colhead{$r_{min}=10R_*$}}
\startdata
0.005& 59.7 & 78.2 & 83.9 \\
0.01& 59.4 & 77.9 & 83.7 \\
0.03& 59.3 & 76.7 & 82.5 \\
0.05& 57.2 & 75.6 & 81.4 \\
0.07& 56.1 & 74.5 & 80.3 \\
0.1& 54.5 & 72.8 & 78.6 \\
0.3& 44.7 & 62.3 & 67.8 \\
0.5& 36.7 & 53.1 & 58.3\\
\enddata
\tablecaption{The limit on the inclination for the star to begin occulting the disk and for the disk to begin occulting the star.}
\end{deluxetable}

\begin{figure}
\epsscale{.5}
\plotone{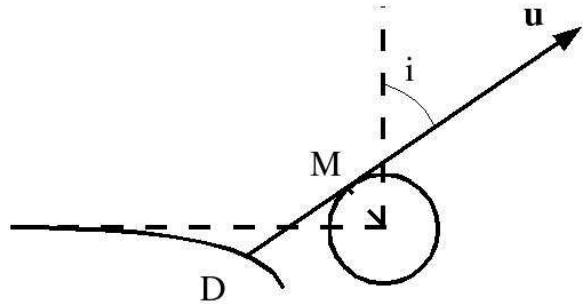}
\caption{Schematic diagram demonstrating the part of the disk that gets blocked by the star. \label{disk_block}}
\end{figure}

If the inclination is greater than $i_{lim}$ then part of the disk is blocked by the star. In this case we can use the discussion of TB96 to determine if the point $P(r,\theta)$ is blocked by the star (figure~\ref{disk_block} and table~\ref{ilim}). Defining $C=r\sin(\theta-\alpha)$ the only time the star can block the disk is when $C<R_*$ otherwise $p(r,\theta)=1$. If $C<R_*$ then $p(r,\theta)=1$ only if $r\cos\theta<r_D\cos\theta_D$, where $\theta_D=\pi-\arcsin(C/r_D)$ and $r_D$ is the positive root of the following equation:

\begin{equation}
\left(-H(r_D)\sqrt{1-\left(\frac{C}{r_D}\right)^2}-z_M\right)\sin i+\left(r_D\sqrt{1-\left(\frac{C}{r_D}\right)^2}+x_M\right)\cos i=0
\end{equation}

where $H(r)=gr_{min}\left(\frac r{r_{min}}\right)^{-n}\cos\alpha$ and M is the point on the northern hemisphere of the star in the plane (P,{\bf u},z) such that {\bf u} is tangent to the star at this point:

\begin{eqnarray}
x_M=-\sqrt{R_*^2-C^2}\cos i\\
y_M=C\\
z_M=\sqrt{R_*^2-C^2}\sin i
\end{eqnarray}

\begin{figure}
\epsscale{.75}
\plotone{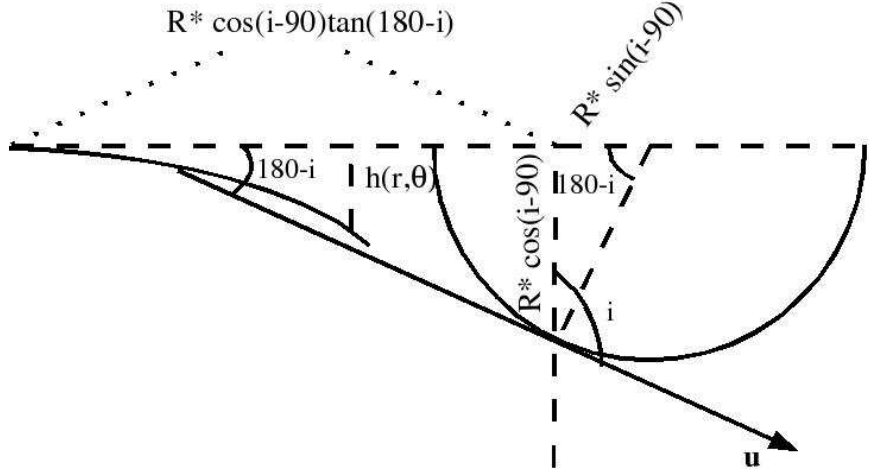}
\caption{Schematic diagram demonstrating the part of the disk that gets blocked by the star. \label{disk_block2}}
\end{figure}

Now consider inclinations greater than $90^{\circ}$. In this case the disk can still be blocked by the star if the warp is small enough and the inclination angle is close enough to 90 degrees (figure~\ref{disk_block2}). This affects both the upper and lower part of the disk, but only from $\theta\in[\alpha+\pi/2,\alpha+3\pi/2]$. The condition for the point P to be hidden by the star is:

\begin{equation}
|h(r,\theta)|<\frac{R_*\cos(i-\pi/2)\tan(\pi-i)-r}{tan(\pi-i)}
\end{equation}

As above this only applies when $C=R_*\sin(\theta-\alpha)<R_*$.

\begin{figure}
\epsscale{.5}
\plotone{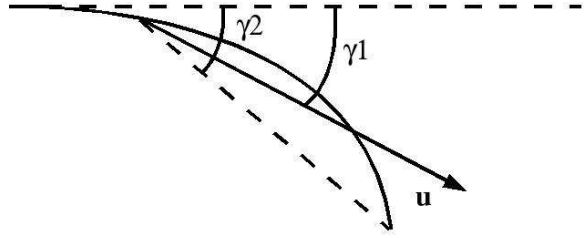}
\caption{Schematic diagram demonstrating when the disk is steep enough to block itself. \label{disk_block3}}
\end{figure}

Another possibility to consider when the inclination is greater than $90^{\circ}$ is that the warp on the lower concave side is steep enough that is blocks part of this side of the disk (figure~\ref{disk_block3}). This condition only applies to parts of the lower concave side, from $\theta\in[\alpha+\pi/2,\alpha+3\pi/2]$. The condition is that $\gamma_1<\gamma_2$ where $\gamma_1=i-\pi/2$ and 

\begin{equation}
\gamma_2=\arcsin\left(\frac{h_m-h(r,\theta)}{PM}\right)
\end{equation}

In this case M is the highest point on the disk on the line of sight to point P. For $r\sin(\pi-\theta)>r_{min}$ the disk does block itself, but for $r\sin(\pi-\theta)<r_{min}$ we have $r_m=r_{min}$ and $\theta_m=\arcsin(r\sin(\theta-\alpha)/r_m)$. The quantity PM is the distance between the point P and the point M ($\sqrt{(h_m-h)^2+(r_m-r)^2}$).

Once all of the conditions have been considered the flux from the disk can be calculated using equation~\ref{eqn_flux}. These conditions would apply to any type of warp whose maximum height above the midplane occurs at the inner edge of the disk, as opposed to at the outer edge of the disk, regardless of the exact functional form of the warp (ie. power law vs. exponential).

\section{Spiral Wave}
\subsection{Temperature Profile of Spiral Wave}
The third type of disk that we attempt to model contains a spiral wave. As with the warped disks we follow the derivation of TB96 to derive the temperature structure and SED for this disk. The derivation for the temperature structure is very similar to that of the inner warp, only with slightly different definitions of the boundaries. For the part of the disk inside the wave, the disk is not blocked by the wave but the amount of the star seen can change. For points far from the wave, the disk is like a flat disk and $\delta_{min}=0$. For points on the wave, as it rises above the midplane, more of the lower half of the star will become visible. How much of the lower half of the star is visible depends on the location and height of the point on the wave. In this case the lower limit on $\delta$ is:

\begin{equation}
\delta_{min}=-\arctan(h/(r-r_{min}))\\
\end{equation}

This limit will continue to increase until the point on the wave can see the entire star and then $\delta_{min}=-\delta_{max}$. The other limits stay the same as in the previous models:

\begin{eqnarray}
\varepsilon_{min}=0\nonumber\\
\varepsilon_{max}=\pi/2\nonumber\\
\\
\delta_{max}=\arcsin(R_*/d)\nonumber\\
d^2=r^2+h^2\nonumber\\
\end{eqnarray}

For the parts of the disk behind the wave, some of the star may be obscured. In this case:

\begin{eqnarray}
\delta_{min}=\arctan(h_{sw}/(r-r_{sw}))\\
h_{sw}=gr_{min-sw}(1-m\theta/2\pi)\nonumber\\
r_{sw}=r_{min-sw}(1+n\theta)\nonumber\\
\end{eqnarray}

This is similar to the concave side of the inner warp, where the warp can obscure part of the star. The only difference is that the maximum height of the disk does not occur at the inner edge of the disk, but instead occurs at the location of the spiral wave. When $\delta_{min}>\delta_{max}$ then the entire star is blocked and that point on the disk is only heated by viscous dissipation. With these definitions and equation (6) we can calculate the temperature of the disk. 

\subsection{Calculating SED of Spiral Wave}
The flux from the disk is given by:

\begin{equation}
F_{\nu,{\bf u}}=\int^{R_{disk}}_{r_{min}}\int^{2\pi}_{0}B(T_{disk})f_{up}
\end{equation}

There is no symmetry in the disk that allows us to split the disk into different parts, as with the concave and convex pieces of the inner warp. We also only consider inclinations less than $90^{\circ}$, where we only see the upper disk, since the lower disk will look the same as the upper disk.

\begin{figure}
\epsscale{0.5}
\plotone{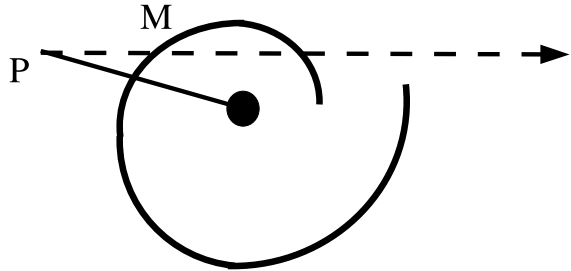}
\caption{Schematic diagram demonstrating the position of M and P in the disk when the wave can block part of the disk.\label{wave_block1}}
\end{figure}

The one occultation effect we include is the blocking of the disk by the wave. For $\alpha-\pi/2<\theta<\alpha+\pi/2$, the near side of the disk, the wave can block the part of the disk that is at smaller radius than the wave. For $\alpha+\pi/2<\theta<\alpha+3\pi/2$, the side of the disk on the other side of the star from the observer, the outer disk can be blocked by the wave. This effect can become important for modest inclinations, given the typical wave heights we consider here. To determine if a point on the disk is blocked by the wave we first need to determine where the line of sight intersects the wave. This is illustrated in figure~\ref{wave_block1} and is given by:

\begin{equation}
x = r\sin\theta = r_{min-sw}(1+n\theta_M)\sin(\theta_M)
\end{equation}

\begin{figure}
\epsscale{0.5}
\plotone{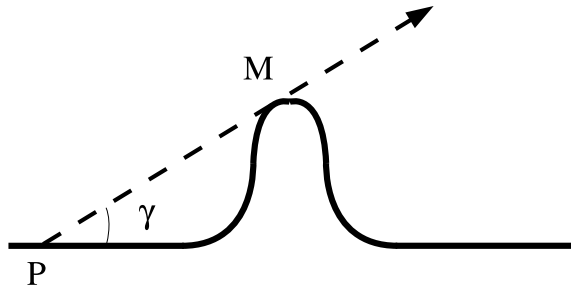}
\caption{Schematic diagram demonstrating how the spiral wave can block part of the disk.\label{wave_block}}
\end{figure}

Here $\theta_m$ is the azimuthal coordinate of the point where the wave intersects the line of sight. We assume that the point M lies at the peak of the spiral wave. This is only an approximation, although the narrowness of the wave make it an accurate one. The angle between the line connecting the points P and M and the midplane is $\gamma$ (Fig.~\ref{wave_block}). When $\gamma>\pi/2-i$ then point $P(r,\theta)$ is blocked.

\begin{equation}
\tan\gamma=\left(\frac{h_M-h_P}{r_M-r_P}\right)
\end{equation}

We ignore occultation effects due to the star blocking the disk, which we did consider in the inner and outer warp model. Based on our experience with the warped disks and the typical dimensions of the disk, these are negligible effects that will only play a role very close to edge on. We also do not consider situations where the wave on the near side of the disk can block the far side of the disk. The exclusion of these two effects prevents us from considering the spiral wave at inclinations very close to $90^{\circ}$.

\section{Stellar Flux}
Next we consider the flux coming from the star. We follow a similar procedure as above where the equation for the stellar flux is modulated by a binary function ($\varepsilon(\phi,\psi)$) which equals 1 when that part of the star is not blocked by the disk and it equals zero when the star is blocked by the disk. The angles $\phi,\psi$, shown in figure~\ref{star}, are the azimuthal and polar angles of a point on the surface of the star relative to the center of the star and the z axis (the same z axis as for the disk). The x-axis of this coordinate system is in the same direction as the line of sight, and will differ from the x-axis of the disk by the angle $\alpha$. The flux from the star is:

\begin{figure}
\epsscale{.25}
\plotone{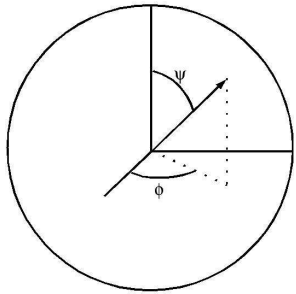}
\caption{Schematic diagram defining the angles $\phi$,$\psi$\label{star}}
\end{figure}

\begin{equation}
F_*=B_{\nu}\int\int_{surface}\varepsilon(\phi,\psi)d{\bf A}
\end{equation}

To determine the surface over which we integrate, we need to know the points of the star that are seen by the observer (ie. which side of the star is facing the observer). These points will be those that have ${\bf u}\cdot d{\bf A}\geq0$ where

\begin{eqnarray}
{\bf u}=\sin i\hat{x}+\cos i\hat{z}\\
d{\bf A}=R_*^2\sin\psi d\psi d\phi(\cos\phi\sin\psi\hat{x}+\sin\phi\cos\psi\hat{y}+\cos\psi\hat{z})
\end{eqnarray}

\begin{figure}
\epsscale{.5}
\plotone{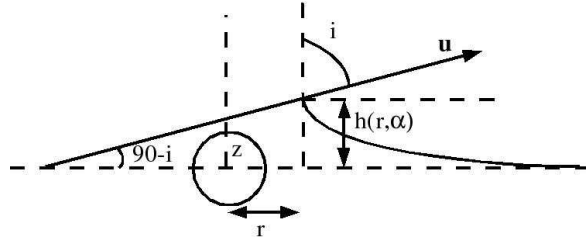}
\caption{Schematic diagram showing the case where $i<90^{\circ}$ and the disk can block the star. \label{star_block}}
\end{figure}

The evaluation of $\varepsilon(\phi,\psi)$ will depend on the type of warp/wave and the orientation of the observer. First consider inclinations less than $90^{\circ}$. In this case the warp/wave streching above the midplane may block some of the star. The entire star will be blocked if the following condition is met:

\begin{equation}
h(r,\alpha)-r\tan(\pi/2-i)>R_*
\end{equation}

where r is location of the peak of the vertical disturbance and $h$ is the maximum height of the warp or wave at the angle $\alpha$. The exact value of $r$ and $h$ will depend on whether we are considering the outer warp, inner warp, or spiral wave (ie. $r=r_{min}$, $h(r,\alpha)=gr_{min}\cos(\alpha)$ for the inner warp). This condition is illustrated in figure~\ref{star_block} for the inner warp, which is the disk that is the most likely to occult the star. 

None of the star will be blocked if the inclination is less than $i_{lim}$ (discussed earlier). A generic version of the equation for $i_{lim}$ that can be applied to all of the disks is:

\begin{equation}
\tan i_{lim}=\frac{r-R_*\cos i_{lim}}{h(r,\alpha)+R_*\sin i_{lim}}
\end{equation}

When the inclination falls between these two limits only a fraction of the star is blocked. We can use the discussion of TB96 section A2.2 to determine if a point of the star's surface is hidden by the star

The point Q is a point on the surface of the star that intersects the line of sight and the upper edge of the disk. If the point $N(\phi,\psi)$ lies above Q then the observer can see this part of the star, otherwise it is hidden and $\varepsilon(\phi,\psi)=0$. The vertical coordinate of Q, $z_Q$, is the greatest root of the following equation:

\begin{equation}
(1+\tan^2 i)z_Q^2-2\tan i(h\tan i-r\cos\alpha)z_Q+(h\tan i-r\cos\alpha)^2-R_*^2+R_*^2\sin^2\psi\cos^2\phi=0
\end{equation}

If $z_N\geq z_Q$ then $\varepsilon(\phi,\psi)=1$ otherwise $\varepsilon(\phi,\psi)=0$ with $z_N$ being given by:

\begin{equation}
z_N=-R_*\sin\psi\sin\phi\sin i+R_*\cos\psi\cos i
\end{equation} 

\begin{figure}
\epsscale{.5}
\plotone{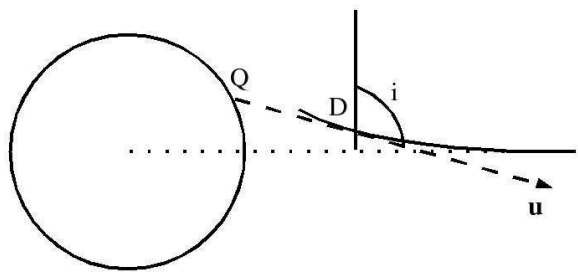}
\caption{Schematic diagram showing the case where $i>90$ and the disk can block the star. \label{star_block2}}
\end{figure}

For the inner and outer warp we consider inclinations greater than $90^{\circ}$ where the disk can still block part of the star, although it is less likely because the disk curves away from the observer. This is illustrated in figure~\ref{star_block2} for the inner warp, but can also apply to the outer warp in the limit that $h$ goes to zero. The point at which the line of sight is perpendicular to the normal of the disk sets a limit to the distance z above the midplane that an observer can see. If this distance is less than the radius of the star then some of the star is blocked by the disk. The point D at which the the line of sight is perpendicular to the disk occurs when ${\bf u}\cdot{\bf n}=0$ (figure~\ref{star_block2}). If this condition is met at a radius $r_D$ then a point $N(\phi,\psi)$ on the stellar surface will be blocked if

\begin{equation}
R_*\cos\psi>h(r_D,\alpha)+r_D\tan(i-\pi/2)
\end{equation}

If there is no point at which ${\bf u}\cdot{\bf n}=0$ then $r_D=r_{min}$ and the same condition for being able to see the star is used. When this condition is met $\varepsilon(\phi,\psi)=0$, otherwise $\varepsilon(\phi,\psi)=1$.

All of these different occultations are combined to determine $\varepsilon(\phi,\psi)$. The flux is determined by integrating over the entire surface and added to the flux from the disk to create the observed SED.


\clearpage

\end{document}